\documentclass[structabstract]{aa}
\usepackage{natbib}
\usepackage{graphicx,color}
\usepackage{txfonts}

\graphicspath{{figuras_paper/}}

\newcommand{\insitu}{\textit{in situ}}
\newcommand{\rmd}{{\mathrm d}}
\newcommand{\tc}{t_{\rm c}}
\newcommand{\Vc}{V_{\rm c}}
\newcommand{\uvec}[1]{ \hat{\bf #1} }

\begin{document}

\title{Dynamical evolution of a magnetic cloud from the Sun to 5.4 AU}
\titlerunning{Evolution of a MC from Sun to 5.4 AU}
\authorrunning{M.S. Nakwacki et al.}

\author{M. S. Nakwacki\inst{1,2} 
\and   S. Dasso\inst{2,3}
\and  P. D\'emoulin\inst{4}
\and C.H. Mandrini\inst{2,5}
\and A.M. Gulisano\inst{2,5}
}

\institute{
Instituto de Astronomia, Geofisica e Ci\^encias Atmosf\'ericas, Universidade de S\~ao Paulo, Brazil 
\email{sole@astro.iag.usp.br}
\and Instituto de Astronom\'ia y F\'isica del Espacio, CONICET-UBA, Argentina
\and Departamento de F\'isica, Facultad de Ciencias Exactas y Naturales, UBA, Argentina
\and Observatoire de Paris, LESIA, UMR 8109 (CNRS), F-92195 Meudon Principal Cedex, France
\and Facultad de Ciencias Exactas y Naturales, FCEN, UBA, Argentina}

\date{Received  / Accepted }
\abstract{
Significant quantities of magnetized plasma are transported 
from the Sun to the interstellar medium via Interplanetary Coronal Mass Ejections (ICMEs).
Magnetic Clouds (MCs) are a particular subset of ICMEs, 
forming large scale magnetic flux ropes. 
Their evolution in the solar wind is complex and mainly determined 
by their own magnetic forces and the interaction with the surrounding solar wind.
}
{In this work we analyze the evolution of a particular MC (observed on March 1998) using \insitu\ observations 
made by two spacecraft approximately aligned with the Sun, the first one at 1 AU from the Sun and the second one at 5.4 AU. 
We study the MC expansion, its consequent decrease of magnetic field intensity and mass density, 
and the possible evolution of the so-called global ideal-MHD invariants.}
{We describe the magnetic configuration of the MC at both spacecraft using different models and compute relevant global quantities 
(magnetic fluxes, helicity and energy) at both helio-distances.
We also track back this structure to the Sun, in order to find out its solar source.
}
{We find that the flux rope is significantly distorted at 5.4 AU.  
However, we are able to analyze the data before the flux rope center is over-passed and compare it with observations at 1 AU.
From the observed decay of magnetic field and mass density, 
we quantify how anisotropic is the expansion, and the consequent deformation 
of the flux rope in favor of a cross section with an aspect ratio at 5.4 AU of $\approx 1.6$ 
(larger in the direction perpendicular to the radial direction from the Sun).  
We quantify the ideal-MHD invariants and magnetic energy at both locations, 
and find that invariants are almost conserved, while the magnetic energy decays as expected with the expansion rate found. 
}
{The use of MHD invariants to link structures at the Sun and the interplanetary medium is supported by the results of this multispacecraft study.
We also conclude that the local dimensionless expansion rate, that is computed from 
the velocity profile observed by a single spacecraft, is very accurate 
for predicting the evolution of flux ropes in the solar wind.}

\maketitle

\section{Introduction} \label{sec:1} 
Coronal Mass Ejections (CMEs) are explosive events that release 
energy in the solar atmosphere. The interplanetary counterparts of CMEs are solar wind (SW) structures known as interplanetary coronal mass ejections (ICMEs). Among them there is a subset, called magnetic clouds (MCs), which exhibit a smooth rotation of the magnetic field direction through a large angle, enhanced magnetic field strength, 
low proton temperature  and a low proton plasma beta, $\beta_p$.
MCs are formed by large scale magnetic flux ropes carrying a large amount of magnetic helicity, magnetic flux and energy away from the Sun. The main characteristics of these structures have been enumerated by \citet{Burlaga80}. 

Several authors have consider MCs as static flux ropes 
\citep[see, e.g.,][]{Goldstein83,Burlaga88,
Lepping90,Burlaga95,Lynch03}. Their magnetic fields 
have been frequently modeled using the Lundquist's model \citep{Lundquist50}, which considers a static and axially-symmetric linear force-free
magnetic configuration.  Many deviations from this model have been also studied: 
e.g., non-linear force-free fields \citep{Farrugia99}, 
non-force free fields \citep{Mulligan99b,Hidalgo02,Cid02}, and several non-cylindrical models \citep{Hu01,Vandas02,Demoulin09c}, all of them being static. 
These models are recurrently used to fit \insitu\ magnetic field measurements within MCs to reconstruct the whole flux rope structure.  These techniques have also been tested by 
replacing the observations by the local values found in a numerical simulation, 
and the output of the models has been compared to the known original full simulation \citep{Riley04}.   
The results of these comparisons show that these \insitu\ techniques can reproduce relatively well the magnetic structures when the spacecraft is crossing the MC near its main axis.

In many cases, MCs present clear characteristics of expansion \citep[e.g.][]{Lepping03b,Lepping08}, 
so several dynamical models have been developed to describe these clouds during their observation time.  
Some of these flux rope models suppose a circular cross-section with only a radial expansion \citep{Farrugia93,Osherovich93b,Farrugia97,Shimazu00,Nakwacki08},
while other models include expansion in both directions, radial and axial
\citep{Shimazu02,Berdichevsky03,Demoulin09,Nakwacki08b}.  A dynamical model with an
elliptical shape was derived by \citet{Hidalgo03}, while a model of 
the expansion with an anisotropic self-similar expansion 
in three orthogonal directions was worked out by \citet{Demoulin08}.

 From single spacecraft observations we cannot directly infer the global structure of
the flux ropes and their evolution through the interplanetary medium 
because they are one single point local measurements.
Several strategies are used to derive more information 
on MCs with multi-spacecraft data, as follows.

With two spacecraft located at a similar distance from the Sun and separated 
by a distance of the order of the cross-section of the encountered flux rope, 
the \insitu\ observations provide data at different parts 
 of the flux rope being negligibly affected by the evolution. 
This is used to test the technique computing the magnetic field in the cross section 
from the data of one spacecraft and/or to have a more accurate reconstruction of the 
magnetic field \citep{Mulligan01,Liu08,Kilpua09,Mostl09}.
 
When the two spacecraft positions are viewed from the Sun with a significant angle,
typically in the interval [10\degr ,80\degr ], one can usually derive an estimation 
of the extension of the flux rope, or at least an estimation of the extension 
of the perturbation (e.g. the front shock) induced by the propagation of the flux rope in the interplanetary medium \citep{Cane97,Mulligan01,Reisenfeld03}.  
A larger number of spacecraft permits to constrain more the evolving magnetic structure, such as in the case analyzed by \citet{Burlaga81}.  
Such studies have been extended to cases where the spacecraft are well separated in solar distance with one spacecraft near Earth and the other one at a few AUs 
\citep{Hammond95,Gosling95,Liu06,Foullon07,Rodriguez08}. 
They show that large MCs/ICMEs have large scale effects in the 
heliosphere (e.g. both at low and high latitudes).
  
When the two spacecraft are separated by spatial scales of the order of one or several AUs, the above analysis should take into account the evolution of the MC with solar distance.
This also implies a more difficult association of the \insitu\ observations at both spacecraft.  
Numerical simulations are then useful tools to check if the events observed at each 
spacecraft are in fact a unique event \citep[e.g., ][]{Riley03}.
Since MCs are moving mostly radially away from the Sun,
the radial alignment (line-up) of two spacecraft is a major opportunity 
to study the radial evolution of a MC, as the MC is crossed at a similar location by the two spacecraft. However, it is not common to find events observed by two nearly radially aligned spacecraft. 
One case was observed by Helios-1,2 close to 1~AU and later by Voyager-1,2 at 2 AU, 
with an angular separation from the Sun of about 10\degr\ (resp. 23\degr ) 
between Helios-1 (resp. Helios-2) and both Voyager-1,2 \citep{Burlaga81,Osherovich93c}. 
A second case was observed by Wind and NEAR spacecraft with an angular separation 
from the Sun of about 1\degr\ and a ratio of solar distances of 1.2 \citep{Mulligan99}. 
A third case was observed by ACE and  NEAR spacecraft with an angular separation from the Sun of about 2\degr\ and a ratio of solar distances of 1.8 \citep{Mulligan01b}. 
Finally, a fourth case was observed by ACE and Ulysses spacecraft with an angular separation from the Sun of about 6\degr\ and a ratio of solar distances of 5.4 \citep{Skoug00,Du07}.  This last case has the advantage of a larger radial separation, 
so that the evolution has a larger effect.  This is the MC selected for a deeper study in this paper.

From the \insitu\ data at different solar distances of the same 
MC one can infer directly the evolution of the magnetic field and plasma quantities.
Such radial evolution is otherwise available only from a statistic analysis of 
a large number of MCs observed individually at various distances 
\citep{liu05,Wang05b,Leitner07,Gulisano10}, with possible bias coming from 
the selection of MCs with different properties.  Another application of line-up 
spacecraft is to derive the evolution of global magnetohydrodynamic (MHD) quantities, 
such as magnetic flux and magnetic helicity \citep{Dasso09b}. They are main quantities to test if the flux rope simply expands or if a significant part reconnects with the SW field.  These global quantities permit also a quantitative link to the related solar event \citep{Mandrini05,Luoni05,Rodriguez08}, and set constraints on the physical mechanism of the associated CME launch \citep{Webb00,Attrill06,Qiu07}.

In this paper we further analyze the evolution of a MC observed at two different helio-distances, 
1 and 5.4 AU \citep{Skoug00,Du07}.  This MC has been selected because this is to our knowledge the best line-up observations of a MC between ACE and Ulysses 
spacecraft. The observations are summarized in Section~\ref{sec:2}. The velocity and magnetic models used to complement the observations are described in Section~\ref{sec:3}.  This spacecraft line-up is an opportunity to follow the evolution of the flux rope and, in particular, the global MHD quantities such as magnetic helicity and flux (Section \ref{sec:4}).   We relate this MC to its solar source in
Section \ref{sec:5}. This complements our understanding of the magnetic field evolution.  
Finally, we discuss our results and conclude in Section \ref{sec:6}.

\section{Observations} \label{sec:2} 
\subsection{Instruments and spacecraft} \label{2.Instruments}

We analyze data sets for SW plasma and magnetic field from ACE and Ulysses spacecraft.
We use the Magnetic Field Experiment \citep[MAG,][]{Smith98} with a temporal cadence of 16 seconds and the Solar Wind Electron Proton Alpha Monitor \citep[SWEPAM,][]{Mccomas98} with a temporal cadence of 64 seconds for ACE spacecraft. For Ulysses spacecraft, we use Vector Helium Magnetometer \citep[VHM,][]{Balogh92} for magnetic field observations with a temporal cadence of 1 second and Solar Wind Observations Over the Poles of the Sun \citep[SWOOPS,][]{Bame92} for plasma observations with a temporal cadence of 4 minutes.
 
When the MC passed through Earth (March 5, 1998) ACE was located at 
$\approx 1$~AU in the ecliptic plane, and in 
a longitude of 164\degr\ in the Solar Ecliptic (SE) coordinate system.
When the cloud was observed by Ulysses (March 25, 1998), 
this spacecraft was located at 5.4 AU from the Sun 
and very near the ecliptic plane, in particular it was 
at a latitude of 2\degr\ and at a longitude of 158\degr , in the SE system.
Thus, the position of both spacecraft differs in 2\degr\ 
for latitude and in 6\degr\ for longitude \citep{Skoug00,Du07}.
This angular separation corresponds to a separation distance 
(perpendicular to the radial direction to the Sun) of $\approx 0.6$ AU at the 
location of Ulysses.
This very good alignment between the Sun and both points of observation
of the same object, gives us a unique opportunity to observe the same MC 
at two different evolution stages in the heliosphere.

\subsection{Coordinate systems} \label{2.Coordinate}

We analyze ACE data in Geocentric Solar Ecliptic (GSE) 
system of reference ($\uvec{x}_{GSE}$, $\uvec{y}_{GSE}$, $\uvec{z}_{GSE}$), 
where $\uvec{x}_{GSE}$ points from the Earth toward the Sun, 
$\uvec{y}_{GSE}$ is in the ecliptic plane and in the direction 
opposite to the planetary motion, and $\uvec{z}_{GSE}$ points to the north pole. 
However, Ulysses data are provided in 
the heliographic Radial Tangential Normal (RTN) system of reference 
($\uvec{R}$, $\uvec{T}$, $\uvec{N}$), in which $\uvec{R}$ points from the Sun to the spacecraft, $\uvec{T}$ is the cross product of the Sun's rotation unit vector ($\uvec{\Omega}$) 
with $\uvec{R}$, and $\uvec{N}$ completes the right-handed system \citep[e.g., ][]{Fraenz02}. 

In order to make an accurate comparison between the observations of vector quantities made from both spacecraft and the orientation of the flux rope at both locations,
we rotate all the vectorial data from Ulysses to the local GSE system of ACE (when this 
spacecraft is at the closest approach distance of the MC axis). 
We describe this transformation of coordinates in Appendix~\ref{sec:appendix}.
This permits to compare magnetic field components in the same frame (Figs.~\ref{fig_GSE_ACE},\ref{fig_GSE_Uly}), 
as well as to compare the orientation of the MC at both positions. 
 
We next define a local system of coordinates linked to the cloud 
\citep[i.e., the cloud frame,][]{Lepping90}
in order to better understand the cloud properties and to compare the results at both positions (such as the axial/azimuthal magnetic flux).  The local axis direction of the MC defines $\uvec{z}_{\rm cloud}$ (with $B_{z,\rm cloud}>0$).
Since the speed of the cloud is mainly in the Sun-Earth direction and
is much larger than the spacecraft speed, which can be supposed to 
be at rest during the cloud observing time, we assume a rectilinear 
spacecraft trajectory in the cloud frame.  The trajectory defines a 
direction $\uvec{d}$; so, we take $\uvec{y}_{\rm cloud}$ in the 
direction $\uvec{z}_{\rm cloud} \times \uvec{d}$ and $\uvec{x}_{\rm 
cloud}$ completes the right-handed orthonormal base ($\uvec{x}_{\rm 
cloud},\uvec{y}_{\rm cloud},\uvec{z}_{\rm cloud}$). Thus, 
{$B_{x, \rm cloud}$, $B_{y, \rm cloud}$, $B_{z,\rm cloud}$} are the 
components of $\vec{B}$ in this new base.

The cloud frame is especially useful when the impact parameter,
$p$ (the minimum distance  from the spacecraft to the cloud axis),
is small compared to the MC radius (called $R$ below).  
In particular, for $p=0$ and a MC described using a cylindrical magnetic configuration, $\vec{B}(r) = B_z(r)
\uvec{z} + B_\phi(r) \uvec{\phi}$, we have $\uvec{x}_{\rm cloud} =
\uvec{r}$ and $\uvec{y}_{\rm cloud} = \uvec{\phi}$ after the spacecraft has crossed the MC axis.
In this case, and for a cylindrical flux rope, the magnetic field data obtained by the spacecraft will show: $B_{x, \rm cloud}=0$, a large and coherent variation of $B_{y, \rm cloud}$ (with a change of sign), and an intermediate and coherent variation of $B_{z, \rm cloud}$, from low values at one cloud edge, 
taking the largest value at its axis and returning to low values at
the other edge ($B_{z, \rm cloud}=0$ is typically taken as the MC boundary). 
 
One possible procedure to estimate the flux rope orientation is the 
classical minimum variance (MV) method applied to the normalized series 
of magnetic field measurements within the estimated boundaries of the MC \citep{Sonnerup67}.   It was extensively used to estimate the orientation of MCs 
\citep[see e.g.,][]{Lepping90,Bothmer98,Farrugia99,Dasso03,Gulisano05} 
and it provides a good orientation estimation when $p$ is small compared to $R$ 
and if the in/out bound magnetic fields are not significantly asymmetric. 
\citet{Gulisano07} have tested the MV using a static cylindrical Lundquist's solution.
They found a deviation of the axis orientation from the model of typically
3\degr\ for $p$ being 30\% of $R$. This deviation remains below 20\degr\
for $p$ as high as 90\% of $R$. 
Another method to find the MC orientation is called Simultaneous Fitting (SF). It minimizes 
a residual function, which takes into account the distance between the observed time
series of the magnetic field and a theoretical expression containing several free parameters, which include the angles for the flux rope orientation and some physical parameters associated with the physical model assumed for the magnetic configuration in the cloud \citep[e.g. ][]{Hidalgo02,Dasso03}.

\section{Modeling the magnetic cloud evolution}\label{sec:3}

\subsection{Self-similar expansion} \label{3.Self}

The evolution of a MC can be described with the model developed 
by \citet{Demoulin08}. In this model, based on previous observations
and theoretical considerations, a few basic hypothesis are introduced.
  Firstly, the MC dynamical evolution is split in two different motions: 
(i) a global one describing the position $\vec{r}_{CM}(t)=D(t) \uvec{v}_{CM}$ 
of the center of mass (CM) with respect to a fixed heliospheric frame 
and (ii) an internal expansion where the elements of fluid are described with respect to the CM frame.
  Secondly, during the spacecraft crossing of the MC, the motion of the MC center is
approximately a uniformly accelerated motion and thus
   \begin{equation} \label{D(t)}
 D(t)=D_0 + V_0 (t-t_0) + a (t-t_0)^2/2\,.
   \end{equation}
  Thirdly, the cloud coordinate system, ($\uvec{x}_{cloud}$, $\uvec{y}_{cloud}$, $\uvec{z}_{cloud}$)
defines the three principal directions of expansion.
  Fourthly, the expansion of the flux rope is self-similar with different expansion 
rates in each of the three cloud main axis. 
In the CM frame, this assumption implies that the position, $\vec{r}(t)$, 
of a element of fluid is described by
 \begin{eqnarray}
  \vec{r}(t) &=& x(t) \, \uvec{x}_{cloud} + y(t) \, \uvec{y}_{cloud} 
               + z(t) \, \uvec{z}_{cloud} \label{xyz_def}\\
             &=& x_0 \, e(t) \, \uvec{x}_{cloud}
              + y_0 \, f(t) \, \uvec{y}_{cloud} + z_0 \, g(t) \, \uvec{z}_{cloud}
                                                     \,, \label{self_similar}
 \end{eqnarray}
where $x(t),y(t),z(t)$ are the fluid coordinates from the CM reference point
at time $t$, and where $x_0,y_0,z_0$ are the position coordinates taken at a reference time $t_0$.  The time functions $e(t)$,  $f(t)$, and $g(t)$, provide the specific time functions for the self-similar evolution. 
Finally, based on observations of different MCs at different distances from the 
Sun \citep[e.g., ][]{liu05,Wang05b,Leitner07,Gulisano10}, we approximate $e(t)$ by the function:
   \begin{equation} \label{e(t)}
   e(t) = (D(t)/D_0)^l \,,
   \end{equation}
and similar expressions for $f(t)$ and $g(t)$, simply replacing the exponent $l$ by $m$ and $n$,
respectively, in order to permit an anisotropic expansion. and proton plasma beta ($\beta_p$).

From the conservation of mass 
we model the decay of the proton density as 
   \begin{equation} \label{np(t)}
   n_{p} = n_{p,0} (D/D_0)^{-(l+m+n)}  \,.
   \end{equation}
From the kinematic self-similar expansion proposed before, 
and assuming an ideal evolution (i.e., non-dissipative, so that the 
magnetic flux across any material surface is conserved), 
the evolution of the magnetic components advected by the fluid is
   \begin{eqnarray}
   B_{x,\rm cloud} &=& B_{x,\rm cloud_0}(D/D_0)^{-(m+n)} \,, \nonumber \\
   B_{y,\rm cloud} &=& B_{y,\rm cloud_0}(D/D_0)^{-(l+n)} \,, \label{bcloud(t)} \\ 
   B_{z,\rm cloud} &=& B_{z,\rm cloud_0}(D/D_0)^{-(l+m)} \,. \nonumber
   \end{eqnarray}

With the above hypothesis and neglecting the evolution of the spacecraft position 
during the MC observation, the observed velocity profile ($V_x$) 
along the direction $\uvec{v}_{CM}$ of the center of mass velocity is expected to be \citep{Demoulin08}:
   \begin{eqnarray}
   V_x &=& -V_0 - a(t-t_0) 
           +V_0 \frac{t-t_0}{D_0/V_0 + t - t_0} \zeta 
                              \label{v(t)full} \\ 
   &\approx & -V_0  + \frac{V_0^2}{D_0} \zeta (t-t_0)     \,, \label{v(t)}
   \end{eqnarray}
where $\gamma$ is the angle between $\uvec{z}_{\rm cloud}$ and $\uvec{v}_{CM}$ and 
   \begin{equation} \label{zeta}
   \zeta= l \sin^2 \gamma + n \cos^2 \gamma  \,.
   \end{equation}
For typical values of MCs we can linearize Eq.~(\ref{v(t)full})
in $t - t_0$ and neglect the acceleration $a$ \citep{Demoulin08}.  This implies that the slope of the observed linear velocity profile provides information on the expansion rate of the flux rope in the two combined directions: 
$\uvec{x}_{cloud}$  and $\uvec{z}_{cloud}$ (since $\zeta$ involves both $l$ and $n$).

\subsection{Magnetic field} \label{3.Magnetic}

Since MCs have low plasma $\beta$ (a state near to a force free field) 
and present flux rope signatures, its magnetic configuration is generally 
modeled using the cylindrical linear force-free field 
$\vec{B^L}=B_0 [J_1(\alpha_0 r) \uvec{\phi} + J_0(\alpha_0 r) \uvec{z}]$ \citep{Lundquist50}.
If the expansion coefficients in the three main cloud axis ($l,m,n$) would be 
significantly different, then an initial Lundquist configuration would 
be strongly deformed. 
However, observations of the MC field configuration are approximately consistent 
with this magnetic configuration
at different helio-distances, ranging from 0.3 to 5 AU \citep[e.g., ][]{Bothmer98,Leitner07}.
So that we expect a small anisotropy on the expansion along different cloud directions 
(i.e., $l \approx m \approx n$). 
From observations of different MCs at significantly different helio-distances 
\citep{wang05,Leitner07} and from observations of the velocity profile slope
from single satellite observations \citep{Demoulin08,Gulisano10}, 
it has been found that $\zeta \approx 0.8$. Since $\zeta$ is a combination
of $l$ and $n$, which depends on $\gamma$ for each cloud, a systematic difference on 
$l$ and $n$ would be detected in a set of MCs with variable $\gamma$ angle. 
Such systematic variation of $l$ and $n$ with $\gamma$ was not found. 

The kinematic self-similar expansion given in the previous section combined 
with a non-dissipative regime, as expected for space plasmas, provide a prediction
for the observed magnetic configuration during the transit of the MC.
Then, assuming an initial Lundquist configuration, the observations are modeled as \citep{Demoulin08}:
  \begin{eqnarray}
   B_{x, \rm cloud}(t) &=& -\frac{p}{\rho(t)} \frac{B_0}{f(t)g(t)} 
                            J_1[U(t)]\,, 
                            \label{bx1}\\
   B_{y, \rm cloud}(t) &=& \frac{\Vc(t-\tc) \sin \gamma}{\rho(t)} \frac{B_0}{e(t)g(t)} 
                            J_1[U(t)]\,, 
                            \label{by1}\\
   B_{z, \rm cloud}(t) &=& \frac{B_0}{e(t)f(t)} 
                            J_0[U(t)]\,,
                            \label{bz1}
   \end{eqnarray}and proton plasma beta ($\beta_p$).
where $U(t)=\frac{\alpha_0 \rho(t)}{\sqrt{e(t)^2+f(t)^2}}$, $\rho(t)=\sqrt{(\Vc (t-\tc) \sin \gamma )^2 +p^2}$, $B_0$ is the strength of the 
magnetic field and $\alpha_0/2$ is the twist of the magnetic field 
lines near the center, at time $t=\tc$. 
By construction of the self-similar expansion, this magnetic field is divergence free
at any time.

Equations (\ref{bx1}-\ref{bz1}) have free parameters which are computed by 
fitting these equations to \insitu\ observations, by minimizing a residual function 
and quantifying the square of the difference between the observed and the predicted value
for the magnetic field components (i.e., a least square fit). 
This provides information on the observed flux rope,
such as its orientation, its extension and its magnetic flux.
We call this method the Expansion Fitting (EF) method, 
and EFI method when isotropy ($l=m=n$) is assumed.
 
\subsection{Global magnetic quantities using the Lundquist's model} \label{3.Invariants_L}

Quantification of global magnetic quantities, such as the so-called ideal-MHD invariants, has been very useful to compare and associate MCs and their solar sources \citep[see, e.g., ][]{Mandrini05,Dasso05c,Dasso09b}.
These MHD invariants are computed using a specific model of the MC
magnetic configuration (Section~\ref{3.Magnetic}). 

For the Lundquist's solution, the axial flux is 
  \begin{equation}  \label{FzL}
  F_{z,Lund} = 2 \pi \int_{0}^{R} B_z \,r \,\rmd r
             = 2 \pi J_1(\alpha_0 R_0) \frac{B_0 R_0}{\alpha_0}  \,,
  \end{equation}
where $R_0$ is the flux rope radius at a reference time $t_0$. 
The azimuthal flux is 
  \begin{equation}  \label{FaL}
  F_{\varphi,Lund} = L \int_{0}^{R} B_\varphi \,\rmd r
                   = (1-J_0(\alpha_0 R_0)) \frac{B_0 L_0}{\alpha_0} \,,
  \end{equation}
where $L_0$ is the axial length of the flux rope at $t_0$. 
For a flux rope staying rooted to the Sun, $L(t)$ is typically 
of the order of the distance to the Sun $D(t)$. 

The relative magnetic helicity is \citep{Dasso03,Nakwacki08}
  \begin{eqnarray}
  H_{Lund} &=& 4 \pi L \int_{0}^{R} A_\varphi B_\varphi \, r \, \rmd r \\
           &=& 2 \pi \left( J_0^2 +J_1^2 - \frac{2 J_0 J_1} {\alpha_0 R_0}  \right)
            \frac{ L_0 B_0^2 R_0^2} {\alpha_0} \,.  \label{HL}
   \end{eqnarray}

The magnetic energy content is not an invariant in MCs. In order to compute
its decay rate we simplify and assume that the MC expansion is isotropic with 
$e(t)=f(t)=g(t)$. 
Then, the magnetic energy ($E_{Lund}$) is computed as \citep{Nakwacki08}
  \begin{eqnarray}  
  E_{Lund}(t) &=& \frac{2 \pi L}{2\mu_{0}} \int_{0}^{R} B^2 \, r \, \rmd r \nonumber \\
              &=& \frac{\pi }{e(t)} 
                  \left( J_0^2 +J_1^2 - \frac{J_0 J_1} {\alpha_0 R_0} \right)
                  \frac{L_0 B_0^2 R_0^2}{\mu_{0}}
               = \frac{E_{Lund}(t_0)}{e(t)}  \,, \label{EL} 
  \end{eqnarray}
where $\mu_{0}$ is the magnetic permeability. Thus, for the isotropic expansion, 
the magnetic energy decays with time as $e(t)^{-1}$. 
In the case of a small anisotropic expansion, a similar decay is expected.

\subsection{Global magnetic quantities using the direct method} \label{3.Invariants_DM}

We define below a method to estimate the global magnetic quantities directly from the observations. 
This method assumes firstly that the cross section is circular, 
secondly that there is symmetry of translation of $\vec{B}$ along 
the main axis of the flux rope, and finally that the impact parameter is low 
(so that, $B_y \approx \pm B_\phi$ and $r \approx x$). 
We also separate the time range covered by the cloud in two branches  and proton plasma beta ($\beta_p$).
\citep[the in-bound/out-bound branches corresponding to the data before/after the 
closest approach distance to the flux rope axis, see {\it e.g.}][]{Dasso05b}, 
and we consider each branch in a separate way.
We call this method DM-in/DM-out, depending on the branch (in-bound/out-bound) that is used.
We define the accumulative magnetic fluxes for the axial and azimuthal field components,  
in the in-bound branch as
 \begin{eqnarray} 
  F_{z,DM-in}(x) &=& 2 \pi \int_{X_{\rm in}}^{x} B_z(x') \, x' \rmd x' \label{Fz_acum} \\ 
  F_{y,DM-in}(x) &=& L \int_{X_{\rm in}}^{x} B_y(x') \, \rmd x'       \label{Fphi_acum} \,,
 \end{eqnarray}
where $X_{\rm in}$ is the $x$ value at the starting point of the flux rope and $\rmd x = V_{x} \, \rmd t$. 
Next, from $B_y(x)$ and $F_{z,DM-in}(x)$ we obtain an expression for the 
magnetic helicity \citep{Dasso06}
 \begin{equation} \label{H_cylind_flux}
  H_{DM-in}  = 2 L \int_{X_{\rm in}}^{X_{\rm center}} F_{z,DM-in}(x) \, B_y (x)\, x \, \rmd x \,.
 \end{equation}

Finally, the magnetic energy is computed from the direct observations
of $B(t)=|\vec{B}(t)|$, 
 \begin{equation} \label{E_DM}
  E_{DM-in}  = \frac{\pi L}{\mu_{0}} \int_{X_{\rm in}}^{X_{\rm center}} B^2(x) \, x \, \rmd x \,.
 \end{equation}

In an equivalent way, we define the same quantities for the out-bound branch.
These equations are used to estimate these global quantities directly 
from \insitu\ observations of the magnetic field, using the components 
of the field and the integration variable ($x$) in the local MC frame. 

\begin{table}
\caption{Timings (dd, hh:mm~UT) of the interfaces for substructures inside the ICME, 
identified with numbered ticks in Figures~\ref{fig_GSE_ACE}-\ref{fig_cloud_Uly}.}
\begin{tabular}{cccc}
\hline
Tick number  & Timing-ACE & Timing-Ulysses & Substructure \\
\hline
1 & 04, 11:00 & 23, 13:30 &             \\
  &           &           & sheath      \\
2 & 04, 14:30 & 23, 22:00 &             \\
  &           &           & MC in-bound  \\
3 & 05, 07:45 & 26, 13:00 &             \\
  &           &           & MC out-bound \\
4 & 05, 20:30 & 28, 00:00 &             \\
  &           &           & back        \\
5 & 06, 02:30 & 28, 09:00 &             \\
\hline
\end{tabular}
\label{table:tabla_ticks}
\end{table}
\begin{figure*}[t!]
\centerline{
\includegraphics[width=\textwidth, clip=]{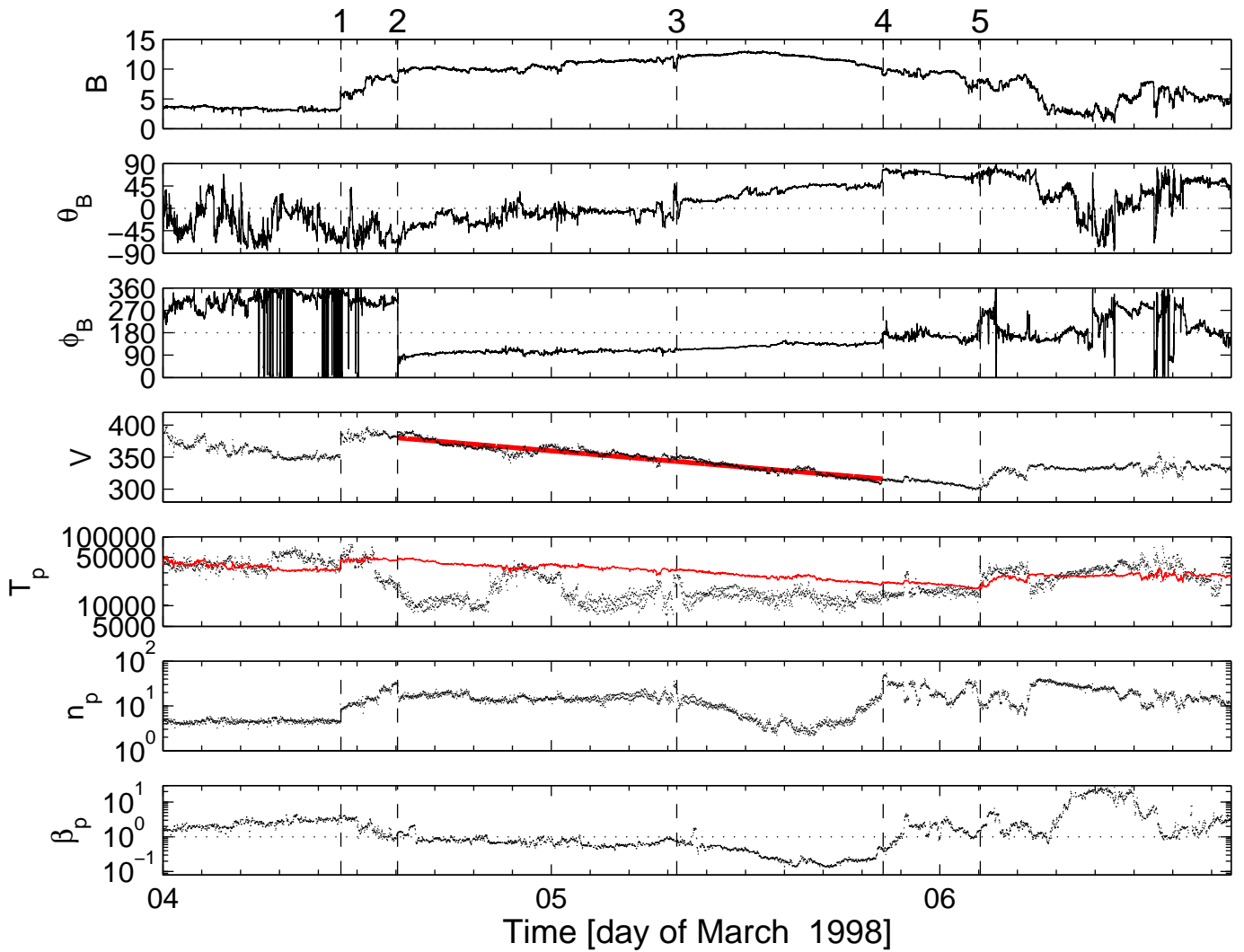}}
\caption{\textit{In situ} plasma and magnetic field of the ICME observed on March 1998 
by ACE located at $\approx 1$ AU from the Sun. From top to bottom: absolute value
of the magnetic field ($B=|\vec{B}|$, in nT), magnetic field vector
orientation (GSE): latitude ($\theta_B$) and longitude ($\phi_B$), bulk
velocity ($V$, in km s$^{-1}$) including in red 
the fitted straight line for the MC range (see Section~\ref{4.Expansion}), 
the expected (continuous red line) and observed (dots) proton temperature
($T_p$, in K), proton density ($n_p$, in cm$^{-3}$), and proton plasma beta ($\beta_p$).
Vertical lines mark different interfaces separating different plasma regions
(see Section~\ref{4.MC_ACE} for a description and
Table~\ref{table:tabla_ticks} for timings). Horizontal dotted lines
in $\theta_B$, $\phi_B$, and $\beta_p$ mark values at 0$^\circ$, 180$^\circ$, 1
as a reference, respectively.}
\label{fig_GSE_ACE}
\end{figure*}

\begin{figure*}[t!]
\centerline{
\includegraphics[width=\textwidth, clip=]{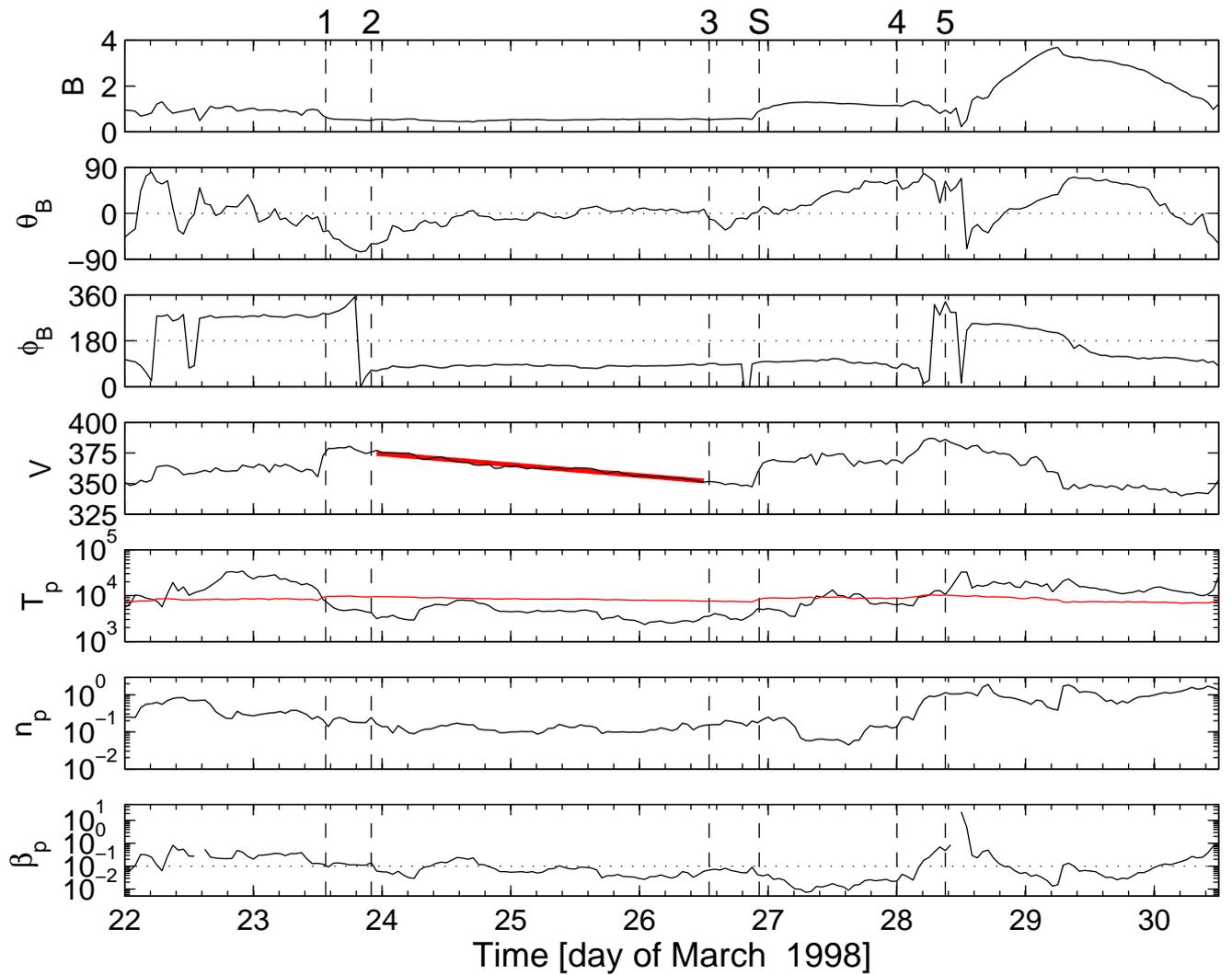}}
\caption{\textit{In situ} plasma and magnetic field parameters of the ICME observed 
on March 1998 by Ulysses located at 5.4 AU from the Sun. 
The format is the same as for Figure~\ref{fig_GSE_ACE}. 
The magnetic field components are in the GSE frame defined at ACE (see Section~\ref{2.Coordinate}). 
Vertical lines with the same reference number correspond to the same interfaces 
of substructures as at 1 AU. 
}
\label{fig_GSE_Uly}
\end{figure*}

\begin{figure*}[t!]
\centerline{
\includegraphics[width=\textwidth, clip=]{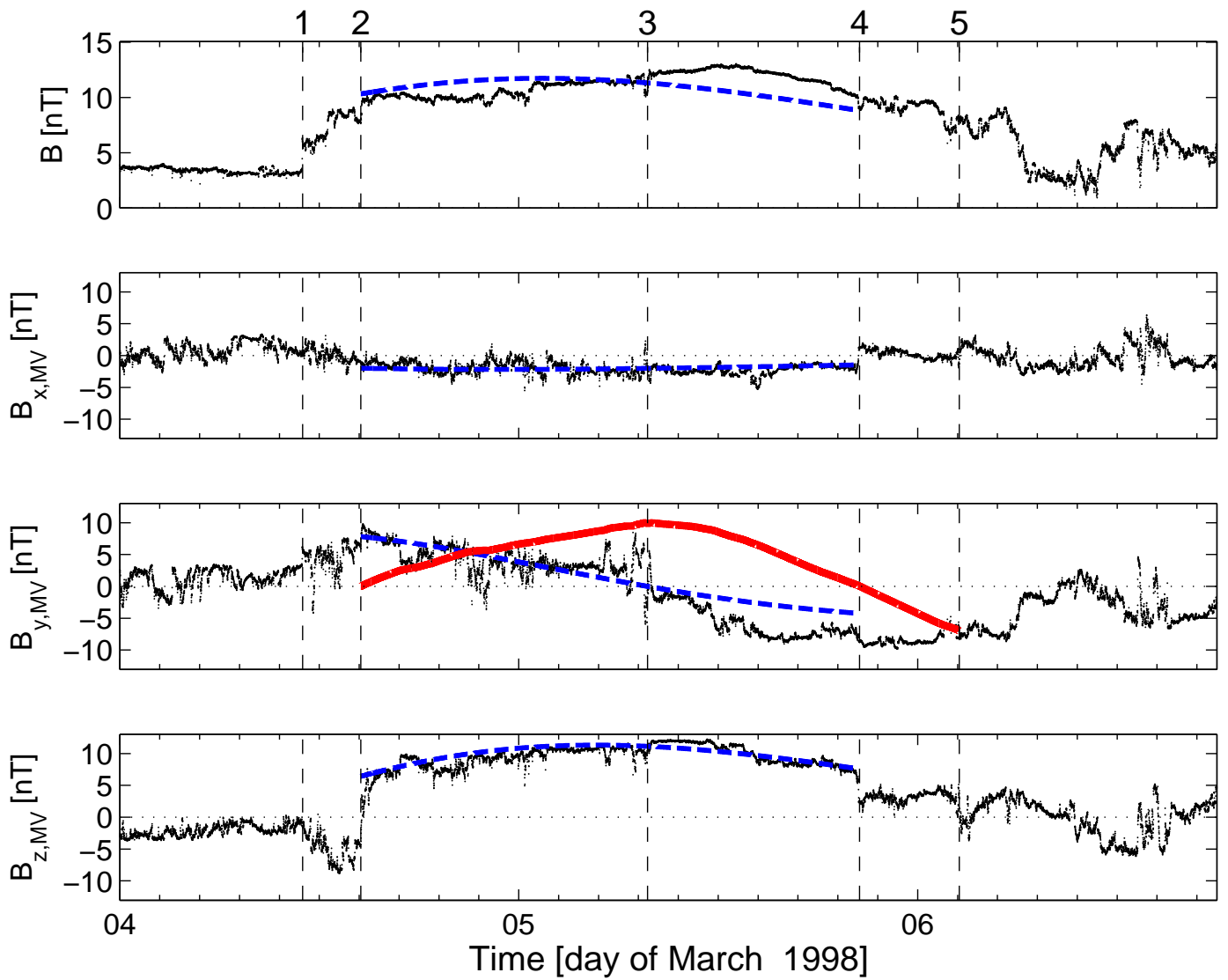}}
\caption{Strength and components of the magnetic field vector 
at 1~AU in the cloud frame given by the MV method (see Section~\ref{4.MC_ACE}).
Numbered ticks are the same as in Figure~\ref{fig_GSE_ACE}.
The dashed blue lines are the Lundquist field model fitted with the EFI method (Section~\ref{3.Magnetic}).
The red thick line is the accumulative magnetic flux for the azimuthal 
component (in arbitrary units).}
\label{fig_cloud_ACE}
\end{figure*}

\begin{figure*}[t!]
\centerline{
\includegraphics[width=\textwidth, clip=]{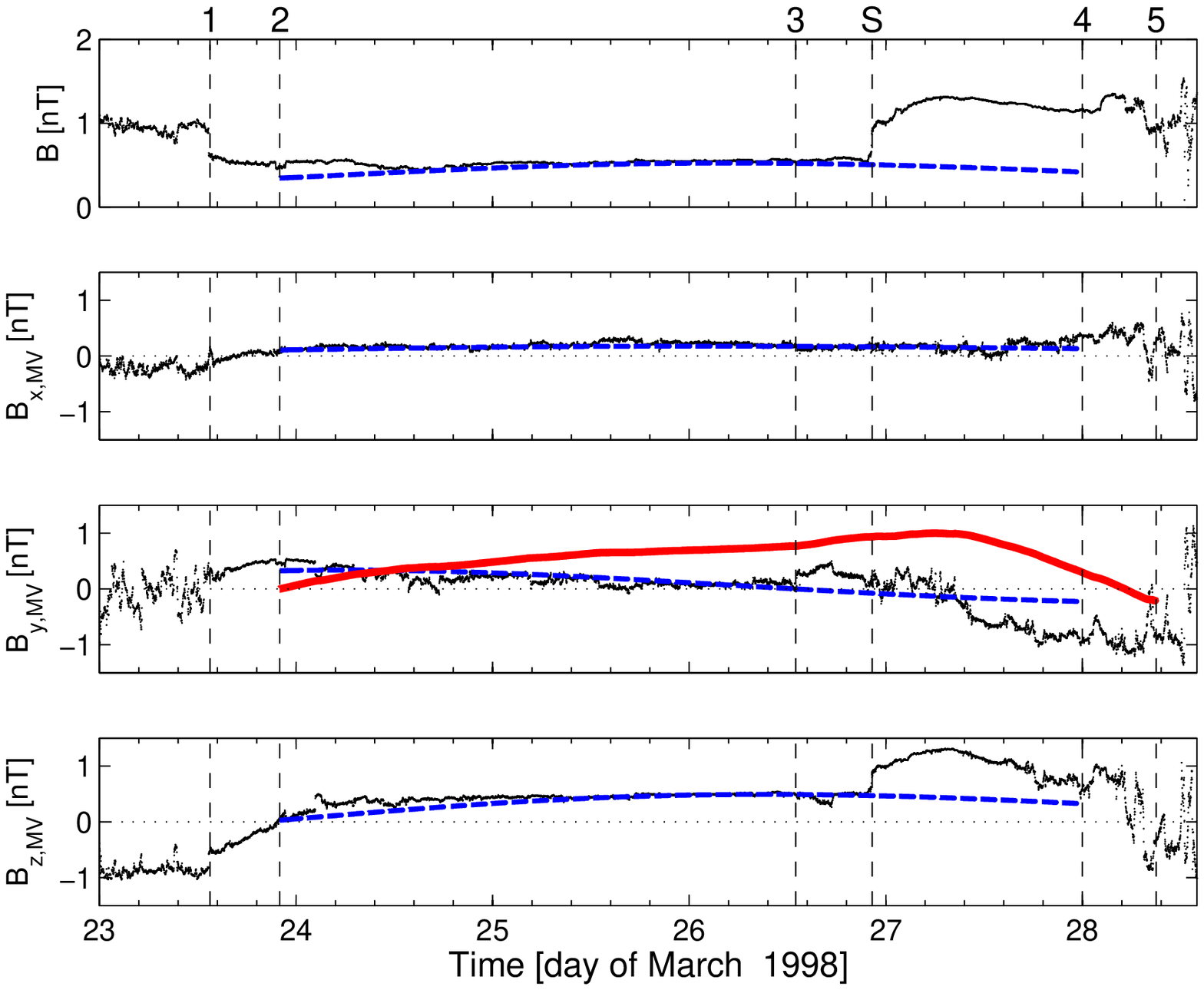}}
\caption{Strength and components of the magnetic field vector 
at 5.4~AU in the cloud frame given by the MV method applied at 1~AU on ACE data 
(see Section~\ref{4.MC_Uly}). The format is the same as for Figure~\ref{fig_cloud_ACE}.
}
\label{fig_cloud_Uly}
\end{figure*}

\section{Results for the studied MC}\label{sec:4}
\subsection{Common frame for data at ACE and Ulysses} \label{4.MC_GSE}

We first describe the \insitu\ data of both spacecraft in GSE coordinates 
(defined at the time of ACE observations, see Appendix~\ref{sec:appendix}).  
Figure~\ref{fig_GSE_ACE} shows the magnetic field and plasma ACE observations. 
The flux rope extends from ticks '2' to '4'. There,
the magnetic field is stronger by a factor $\approx 3$ than in the surrounding SW 
and it has a coherent rotation.  At ACE position the velocity 
profile is almost linear within the MC (Figure~\ref{fig_GSE_ACE}).
The density is relatively important, up to $\approx 20$~cm$^{-3}$, compared to the 
density present in the SW before the MC sheath, $\approx 5$~cm$^{-3}$.  
In most of the MC the proton temperature is clearly lower than the 
expected temperature (red line) in a mean SW with the same speed 
\citep{Lopez86, Elliott05,Demoulin09b}; this is a classical property of MCs. 

Figure~\ref{fig_GSE_Uly} shows Ulysses observations, in the same format as Figure~\ref{fig_GSE_ACE}.
At Ulysses position, the linear velocity profile is still present before a strong shock 
on March 26, 22:20 UT (tick 'S' in Figure~\ref{fig_GSE_Uly}).
The proton temperature is, as typically observed in MCs, well below
the temperature expected for a typical SW at 5.4 AU from an extrapolation of the empirical law given by \citet{Lopez86}. 
Conversely to ACE, the density is much lower in the MC than in the SW 
present before the MC sheath ($\approx 0.1-0.2$~cm$^{-3}$ 
in the MC compared to $\approx 0.5-1$~cm$^{-3}$ in the SW). With mass
conservation, this implies that the volume expansion rate of the MC is much higher than 
the SW one.  The magnetic field observed at Ulysses, has very significantly decreased with respect to the field at ACE (factor $\approx 20$), becoming even weaker than the 
one present in the surrounding SW (typically by a factor $\approx 2$).  This is 
consistent with the important observed decrease in density.

\begin{table}
\caption{Magnetic cloud orientation according to minimum variance and simultaneous fitting.}   
\label{table:orientation}     
\centering  
\begin{tabular}[h]{ l  c  c  c }
\hline
Model & Parameter & ACE & Ulysses\\
\hline
Min. Var & $\theta $ &  ~12\degr &  -29\degr \\
Sim. Fit & $\theta $ &  -11\degr &  -14\degr \\
\hline
Min. Var & $\phi $   &  101\degr & ~~84\degr \\
Sim. Fit & $\phi $   &  115\degr & ~~81\degr \\
\hline
\end{tabular}
\end{table}

\subsection{Magnetic cloud at ACE} \label{4.MC_ACE}

The definition of the MC boundaries is an important step in 
the analysis of a MC since the selected boundaries are affecting 
all the physical quantities related to the magnetic field.
The MC boundaries are associated to discontinuities in the magnetic 
field because such discontinuities are formed in general at the boundary 
of two regions having different magnetic connectivities, such as 
the flux rope and its surrounding medium magnetically linked to the SW. 

The front boundary of MCs is typically better defined than the back (or rear) boundary.  Such is the case in the analyzed MC at ACE, where there is a fast forward shock at tick '1', and a strong discontinuity of $\phi_B$ (at tick '2' in Figure~\ref{fig_GSE_ACE}). 
Moreover, the magnetic field in front has a reverse $\phi_B$ 
and, between ticks '1' and '2', $\vec{B}$ is fluctuating, a characteristic of MC sheaths.
The density has also a discontinuity and is enhanced just before tick '2', 
another characteristic of MC sheaths.  Then, the MC front boundary is set at tick '2'. 
It is worth noting that the shock at tick '1' was previously identified in Wind spacecraft data at 11:05 UT as an ICME related shock 
(see http://pwg.gsfc.nasa.gov/wind/current\_listIPS.htm).
This interplanetary fast forward shock-wave has been recently studied in a multispacecraft analysis by \citet{Koval10}. 
For further details on the characteristics of shocks and their identification in the interplanetary medium see \citet{Vinas86} and \citet{Berdichevsky00}.

The back boundary is also set at a discontinuity of the magnetic field.
As typical in MCs, there are several possibilities after 18 UT 
on March 5 (see Figure~\ref{fig_GSE_ACE}).  
However, $\phi_B$ and $\theta_B$ have clear discontinuities, similar 
but weaker than the front one, at tick '4'. This is confirmed by 
a discontinuity in the density.  The above boundaries are used to find 
the orientation of the flux rope both with the MV and the SF (Section~\ref{2.Coordinate}).  
Indeed, the MC boundaries are better defined in the MC frame since 
the axial and azimuthal field components are separated (Figure~\ref{fig_cloud_ACE}). 
So we need to determine the MC frame.

We first use the MV method to find the direction of the MC axis.
The MV should be applied only to the flux rope, otherwise if part of the back 
is taken into account the directions given by the MV could be significantly 
bias \citep[the back region is no longer part of the flux rope at the observation time,][]{Gulisano07}.  Then, an iteration is needed
starting with the first estimation of the flux-rope boundaries from the data, 
performing the MV analysis, then plotting the magnetic field in the cloud coordinates, 
and finally checking if the selected back boundary is correct with the accumulated 
azimuthal flux (Eq.~(\ref{Fphi_acum})).  
For the studied MC at ACE, the orientation found with the MV provides 
an almost vanishing accumulated flux at the back boundary selected above (Figure~\ref{fig_cloud_ACE}).  Then, no iteration is needed, and the MV is providing a trustable orientation 
within the accuracy of the method (Table~\ref{table:orientation}).
The small mean value of $B_{x,cloud}$ indicates that the impact parameter is small (Figure~\ref{fig_cloud_ACE}).
This implies that the orientation found by the MV method could differ from the 
real one by typically $10$\degr\  \citep{Gulisano07}.

  The MC axis direction is also estimated with a standard simultaneous fit (SF) of 
the Lundquist's solution to the observations (see end of Section~\ref{2.Coordinate}). 
The fitting minimizes the distance of the model to the observed magnetic field in 
a least square manner.  As for the MV, it is important to take into account only the 
data in the range where the flux rope is present. The output of the SF method provides 
estimations for the MC frame vectors as the MV, and also the impact parameter and 
the physical parameters (the free parameters of the Lundquist's solution).
There is a significant difference in the MC axis orientations between the MV and the SF methods, larger in latitude (23\degr ) than in longitude (14\degr ). 
Moreover, with the SF results the accumulated azimuthal flux is significant at the back boundary (so the magnetic flux is not balanced), and there are no nearby significant discontinuities. 
We conclude that the MV method is providing a better approximation of the MC frame than the SF method in this particular MC.
Since $\phi $ is close to 90\degr , it is a good assumption to consider that ACE crosses 
the flux rope front (or nose). 
Moreover since $\theta$ is small, the MC axis lies almost in the ecliptic plane.

In summary, at ACE we select the MC boundaries as March 4, 1998, at 14:30 UT and 
March 5, 1998, at 20:30. The front boundary differs by up to $\approx 5$h from previous studies. 
The larger difference is with the back boundary of the MC since it was set 
on March 06 at 03:15 \citep{Skoug00}, 2:30 UT \citep{liu05}, and 06:30 UT \citep{Du07}, 
so significantly later than our boundary set at tick '4'.
Our boundaries take into account the extension of the flux rope when it was crossing ACE. 
In fact part of the MC characteristics are present after the back boundary '4' 
(e.g. strong and relatively coherent magnetic field, cooler temperature than expected), 
but some other quantities are closer to typical SW values such as $\beta_{p}$ 
which is larger than $1$.
This was previously found in other MCs \citep{Dasso06,Dasso07}, and was called a back region.
This type of region was interpreted as formed by a magnetic field and plasma 
belonging to the flux rope when it was close to the Sun, and later connected to the 
SW due to magnetic reconnection in the front of the flux rope, which in a 
low density plasma as the SW could be more efficient due to 
the Hall effect \citep[e.g.,][]{Morales05}.   So a back region has 
intermediate properties between MC and SW since it is a mixture of the two.  Indeed,
this is the case at the time of ACE observations since $B_{z,\rm cloud}$ 
is fluctuating in the back region, while $B_{y,\rm cloud}$ retains more its coherency 
as in the flux rope (Figure~\ref{fig_cloud_ACE}).

\subsection{Magnetic cloud at Ulysses} \label{4.MC_Uly}

At Ulysses the MC has a complex structure in most of the measured parameters. 
The proton temperature is significantly below the expected temperature 
in the interval of time between tick '1' and about one day after tick '3'  
(Figure~\ref{fig_GSE_Uly}).  At the beginning of this time interval, 
the magnetic field has a coherent rotation, then later it is nearly 
constant up to a large discontinuity of the field strength 
(at tick 'S', $\approx 0.4$ days after tick '3'). 
Later, the magnetic field is stronger with a significant rotation.
Then, the observations at Ulysses show a flux rope with characteristics significantly 
different from a ``standard'' flux rope such as observed at ACE.  
A magnetic field strength flatter than at ACE is expected if the 
flux rope transverse size increases much less rapidly than its length,
since it implies that the axial component becomes dominant with increasing 
distance to the Sun so that the magnetic tension becomes relatively weaker 
in the force balance \citep{Demoulin09}. The strong discontinuity followed by a strong 
field is more peculiar. Indeed, this MC was strongly overtaken by a faster structure, 
as shown by the velocity panel of Figure~\ref{fig_GSE_Uly}, and previously 
identified by \citet{Skoug00}.

The proton temperature has a similar variation at both spacecraft since 
it is becoming lower than the expected temperature approximately after 
the shock defined by the discontinuity of $V$. This discontinuity defines 
the position of tick '1' (Figure~\ref{fig_GSE_Uly}).
The front boundary of the MC is less well defined than at ACE since 
there is no discontinuity.  Still the behavior of the magnetic field 
is similar at both spacecraft as follows.  
Before tick '2', $\theta_B$ is fluctuating while globally decreasing, 
while after tick '2' it is gradually increasing (with fluctuations) at both spacecraft.
$\phi_B$ has a global behavior similar to a step function at both spacecraft, 
with nearly constant values both before tick '1' and after tick '2'.
The main difference for $\phi_B$ is a discontinuity at tick '2' at 1~AU, 
while $\phi_B$ has a smooth transition at 5.4~AU.
Then, we fix the beginning of the MC, tick '2', at the beginning of the period 
where $\theta_B$ starts increasing and $\phi_B$ is nearly constant.  
This is confirmed in the MC frame, 
as $B_{z,\rm MV} \approx 0$ at tick '2' at both 1 and 5.4 AU 
(Figs.~\ref{fig_cloud_ACE},\ref{fig_cloud_Uly}).

The back boundary is much more difficult to define since there are several
possibilities, and the absence of a clear coherence between 
all the measured parameters. 
We tried several back boundaries, the MV and the SF methods, 
and use the iterative method described in Section~\ref{4.MC_ACE} 
to check the selected boundary in the derived MC cloud frame.  
We found that the MC axis orientation can change by more than 20\degr\  
and that the selected boundaries are not confirmed 
by the cancelation of the accumulated flux. 

The difficulties of both the MV and the SF methods are due to the 
large asymmetry of the MC when observed at Ulysses.  
Even normalizing the magnetic field strength is not sufficient since 
the out-bound branch is strongly distorted after the shock (tick 'S').  
This is the consequence of the overtaking flow seen after tick '5' 
in Figure~\ref{fig_GSE_Uly}.
The MC observations have characteristics comparable to the MHD simulations 
of \citet{Xiong07,Xiong09} where they modeled the interaction of a flux 
rope with another faster one.  At Ulysses, the internal shock has propagated 
nearly up to the MC center, so that most of the out bound branch is strongly distorted.
This implies that both the MV and the SF methods cannot 
be used to find the MC orientation at Ulysses.

In the MC frame deduced at ACE, the magnetic field components measured 
at Ulysses between ticks '2' and '3' have the expected behavior for a flux rope 
(Figure~\ref{fig_cloud_Uly}),
as follows.  $B_{x,\rm cloud}$ is almost constant and small indicating 
a low impact parameter. $B_{y,\rm cloud}$ shows a clear rotation 
and $B_{z,\rm cloud}$ is increasing from tick '2' to '3'. These are 
indications that the studied MC has not significantly changed its 
orientation ($\leq$ 10\degr ) between 1 and 5.4 AU, despite the presence of the SW 
overtaking the rear region of the 
flux rope.
Therefore, for Ulysses we use below the MC frame found at ACE.

The accumulated azimuthal flux is maximum after the 
internal shock position (Figure~\ref{fig_cloud_Uly}), an indication 
that the shock would have overtaken the flux rope center.
However, we cannot trust the behavior of the accumulated azimuthal 
flux behind the shock since the orientation and the strength of 
the magnetic field are strongly modified.  Indeed, all the magnetic 
field components are perturbed even before the shock, in the interval ['3','S'].  
Before tick '3', $B_{y,\rm cloud}$ is weak and it almost vanishes at 
tick '3', while $B_{z,\rm cloud}$ is nearly maximum there, 
so we set the flux rope center at tick '3'.

In summary, at Ulysses we initially select the ICME boundaries, 
including the sheath, as from March 23 at 13:30 UT (tick '1') 
to March 28 at 09:00 UT (tick '5'). 
The boundaries chosen by \citet{Skoug00} were slightly different since
they chose the range from March 24 at 02:00 UT to March 28 at 02:30 UT.
The above detailed analysis of the magnetic field behavior indicates 
that the in-bound branch of the MC is from tick '2' to '3', with an out-bound 
branch from '3' to '4' strongly distorted. 
The in-bound extension is confirmed by a similar behavior of the 
velocity and the proton temperature at 1 and 5.4 AU.

\subsection{Impact parameter at ACE and Ulysses} \label{4.Geometrical}
The agreement between the characteristics of the MC observed at ACE and Ulysses
are clear indications that the same MC was observed. 
In this section we estimate the impact parameter when the MC is observed at each of the two spacecraft.
 
Figure~\ref{fig_cloud_ACE} shows that $B_{x,cloud}$ is small compared to $B$, 
as expected when the impact parameter ($p$) is close to zero. 
From the mean value of $B_{x,cloud}$, we estimate $p/R \approx 0.3$ \citep[see the method in][]{Gulisano07}. From the in-bound size $S_{ACE,in}$ =0.144 AU and $p/R=0.3$, we estimate that the radius $R_{ACE} \approx 0.15$ AU, with a circular cross-section.
Fitting the value of $p/R$, using the SF-EFI method, we also obtain $p/R \approx 0.3$.
Another estimation of the impact parameter can be done using the approximation
introduced in the simplest form of Eq.~(31), 
$<B_{x,cloud}> / <B>  \approx 1.2 p/R$ \citep{Demoulin09c}, assuming a cross section roughly circular.  We obtain $p/R \approx 0.27$ from this method.  Thus, in conclusion, the impact parameter when ACE observes the cloud is low ($p/R \approx 0.3$) and very similar from its estimation using different proxies.

From the mean value of $B_{x,cloud}$ at Ulysses (see Figure~\ref{fig_cloud_Uly}), as done for ACE, we estimate $p/R=0.54$.
From $S_{Uly,in}$ =0.55 AU and $p/R=0.54$, we estimate that $R_{Uly} \approx 0.65$ AU. The above values of $R$ at both spacecraft are in agreement with the values found from fitting the free parameters of the cylindrical expansion Lundquist 
model (EFI method, see Section~\ref{3.Magnetic}, Eqs.~(\ref{bx1}-\ref{bz1})).
Again, using also the estimation of the impact parameter from the simplest approximation
of \citet{Demoulin09c} ($<B_{x,cloud}> / <B>  \approx 1.2 p/R$),
we obtain from this method that $p/R \approx 0.18$. 
Thus, in conclusion, the impact parameter when Ulysses observes the cloud is also low, with $p/R$ in the range [0.2-0.5] using different proxies.

Further arguments in favor of the association of the MC observed at ACE and Ulysses are given in the following two sections with the timing and the agreement of the mean velocity, expansion rate, magnetic fluxes and helicity as deduced at the two locations.

\subsection{Acceleration and Expansion} \label{4.Expansion}

The translation velocity of the flux rope at ACE is estimated, as typically done, 
as the mean value of the observed bulk speed during the observation range ['2','4']. It turns out to be $V_{\rm ACE}=-348$~km/s.
However, for Ulysses, because of the perturbation on the out-bound branch,
it is not possible to apply this classical procedure. 
Then, we compute a mean value of the observed speed only in a symmetric range near 
the center (tick '3'). We choose a range of 12 hours around '3', which gives $V_{\rm Ulysses}=-351$~km/s. 
So that at Ulysses, the MC travels slightly faster than at ACE, with a mean acceleration 
of $<\!a\!> = |V_{\rm Ulysses}-V_{\rm ACE}|/\Delta t=0.14~$km/s/d, 
where $\Delta t=21.2$ days is the elapsed time between both centers. 
However, the small value obtained for the acceleration just indicates that it was almost negligible
during the transit from ACE to Ulysses and it implies a negligible contribution 
to the interpretation of the observed velocity profile as a proxy of the expansion of the flux rope \citep[Eq.~(\ref{v(t)full}), see][for a justification]{Demoulin08}. 
Then, we consider below that the MC is traveling from ACE to Ulysses with a constant velocity.

We fit the velocity observations to the velocity model described in Sect.~\ref{3.Self} (Eq.~\ref{v(t)}).  From the fitted curve (red line, panel 4 Fig.~\ref{fig_GSE_ACE} and Fig.~\ref{fig_GSE_Uly})
we obtain the expansion coefficients $\zeta_{ACE} = 0.74$ for ACE and $\zeta_{Uly} = 0.67$ for Ulysses.
These values are very similar at both helio-distances
and they are consistent with previous results found at 1~AU with the
same method for a set of 26~MCs \citep{Demoulin08}, as well as the results obtained
from a statistical analysis of MCs or ICMEs observed at various distances of the Sun
\citep{Bothmer94,chen96,liu05,Leitner07}.

The angle, $\gamma$, between the axis of the MC and the direction of motion 
defines the contribution of the axial and ortho-radial expansion rate to the 
observed $v_x(t)$, see Eq.~(\ref{zeta}). From the orientation given by MV at ACE, this angle is $\gamma = 101\degr $.
This implies that the expansion rate measured from $v_x$ is mainly due to the expansion in a direction perpendicular to the cloud axis and, then, for the cloud
analyzed here $\zeta \approx l$. 

The value of $\zeta$ observed at a given spacecraft, just provides 
the 'local' expansion rate, which corresponds to the expansion during the \insitu\ observations.  However, since $\zeta_{ACE} \approx \zeta_{Uly}$, we assume that during the full travel between ACE and Ulysses the mean expansion rate occurred at a value between $\zeta=0.67$ and $\zeta=0.74$.

Assuming a self-similar expansion in the Sun-spacecraft direction 
(as in Eq.~(\ref{e(t)})), we can link (without any assumptions on the cloud shape and symmetries for the flux rope) the size of the structure along the direction 
of the MC motion ($\uvec{v}_{CM}$), at both helio-distances with the 
expected size at Ulysses given by $S_{ULY,exp}=S_{ACE,obs}(5.4)^\zeta$.

Because the out-bound branch of the MC at Ulysses is strongly distorted, 
we analyze the in-bound branch. 
Then, using the mean velocity and the time duration between ticks '2' and '3', 
we find that $S_{ACE,in}=$0.144 AU. Then, from the observed value of $\zeta_{ACE}$, 
we find an expected size at Ulysses of $S_{Uly,exp-in}=$0.498 AU 
(using the observed $\zeta_{Uly}$, we obtain $S'_{Uly,exp-in}=$0.444 AU).
This implies an expected center time on March 26, 9:00 UT 
(and even an earlier time when using $\zeta_{Uly}$). However, 
the cumulative azimuthal flux ($Fy$, Figure~\ref{fig_cloud_Uly} panel~3) is smoothly and monotonically increasing beyond this time, 
with a strong discontinuity of $By$ a bit later, on March 26, 13:00 UT.  
At the center of the flux rope we expect a global extreme of $Fy$ \citep{Dasso06,Dasso07,Gulisano10}, 
and, in particular, as it was observed when this same cloud was located 
at 1AU (panel 3 of Figure~\ref{fig_cloud_ACE}). 
However, it is not the case in Ulysses at the expected 
time (9:00 UT), so that we decide to take the center at 13:00 UT,
the time when $B_{y}$ started to be disturbed.
This lack of exact agreement between the predicted and observed center positions could be associated with the not exact alignment between ACE and Ulysses.

\begin{table}
\caption{
Observed (ACE and Ulysses) mean values of proton density ($n_p$) and 
magnetic field ($B$, $B_{y,MV}$, and $B_{z,MV}$) for the in-bound 
branch of the magnetic cloud, for three different estimations of expansion rates (see Section~\ref{4.Prediction}).
The predicted to observed ratio, at Ulysses, is shown in parenthesis.}
\label{table:Pred_obs}
\centering
\begin{tabular}[h]{  l  l  l  l  l l l l l} 
\hline
Quantity & ACE & Uly & UlyP1 & UlyP2 & UlyP3\\
\hline
l &    &   & 0.7 & 0.7 & 0.74 \\
m &    &   & 0.7 & 1. & 1. \\
\hline
$np$ (cm$^{-3}$) & 15.8 & 0.12  & 0.27 (2.2) & 0.17 (1.4) & 0.16 (1.3) \\
$B$ (nT)         & 10.3 & 0.48  & 0.88 (1.7) & 0.56 (1.1) & 0.53 (1.0) \\
$B_{y,MV}$ (nT)  &  4.1 & 0.18  & 0.51 (2.8) & 0.24 (1.3) & 0.22 (1.2) \\
$B_{z,MV}$ (nT)  &  9.0 & 0.42  & 0.85 (2.0) & 0.51 (1.2) & 0.48 (1.2) \\
\hline
\end{tabular}
\end{table}

\subsection{Prediction of the mean plasma density and magnetic field at Ulysses} \label{4.Prediction}

The expected values of the proton density and magnetic field at Ulysses can be predicted using the observations at ACE and the expansion rates along the three directions ($l$, $m$, and $n$, see Eqs.~(\ref{np(t)}-\ref{bcloud(t)})).
The mean expansion rate, $\zeta$, along the plasma flow can be estimated with a value between $\zeta_{ACE}$ and $\zeta_{Uly}$, 
which for the orientation of this cloud results to be $l \approx \zeta$. 
The presence of bi-directional electrons supports the 
connectivity of this MC to the Sun \citep{Skoug00}; then, 
the axial expansion rate is estimated as $n\approx 1$, 
because the axial length needs to evolve as $D$ in order to keep the magnetic 
connectivity of the MC to the Sun \citep[e.g.,][]{Demoulin09}.
For the third expansion rate, $m$, we have no direct observational constraint.

Because after 'S' at Ulysses (see Figure~\ref{fig_GSE_Uly}) the cloud is strongly perturbed, we compare ACE and Ulysses in the in-bound branch.
Column 2 of Table~\ref{table:Pred_obs} shows the mean values inside the in-bound branch, for the proton density ($n_p$) and the magnetic field observed at ACE. 
Column 3 shows the same mean values, but now observed at Ulysses. Columns 4-6 show three predictions for these quantities at Ulysses.
The field strength, $B$, is computed from $B_{y,MV}$ and $B_{z,MV}$, since $B_{x,MV}$ is dependent on the impact parameter which differs at both spacecraft (anyway $|B_{x,MV}|<<B$ so that including $B_{x,MV}$ does not change significantly the results). 
All the predictions are made using $n=1$.  Column 4 (UlyP1) shows the prediction at Ulysses using 
$l = (\zeta_{ACE}+\zeta_{Uly})/2 = 0.7$  and $m=l$  (i.e., an isotropic expansion in the plane perpendicular to the cloud axis).
Column 5 (UlyP2) shows the prediction at Ulysses using $l=0.7$ and $m=1$, which corresponds to an expansion such that the cross section of the magnetic cloud is deformed toward an oblate shape in the plane perpendicular to the cloud axis, with the major axis perpendicular to the global flow speed \citep[e.g., ][]{Demoulin09c}.
Column 6 (UlyP3) shows the prediction at Ulysses using $l=0.74$ and $m=1$, which corresponds to an expansion in the direction of the plasma main flow as observed at ACE, 
emulating the case in which the expansion in this direction was similar to that observed 
at ACE almost all the time.

The assumption $l=0.70$ and $m=l$ (UlyP1) predicts significantly larger values for all the quantities with respect to the observed ones (Table~\ref{table:Pred_obs}).  However, the assumption $l=0.70$ and $m=1$ (UlyP2)  provides more realistic predictions.   Furthermore, the assumption $m=1$, combined with $l=0.74$, gives predictions closer to the observations.
Of course due to the lack of a perfect alignment we do not expect an exact matching even when the expansion is well modeled.

Another possible approach is to compute $l,m,n$ from the observed ratio (Ulysses with respect to ACE observations) 
of the mean values of $n_{p}$, $B_{y,MV}$, and $B_{z,MV}$ in the in-bound.  We find $l=0.78$, $m=1.04$, and $n=1.06$.
The value of $l$ is close to $\zeta_{ACE}$, measured independently from \insitu\ velocity, and $n$ is
close to the expected value obtained with a flux-rope length proportional to the distance to the Sun.

We conclude that the expansion of this cloud between ACE and Ulysses was such that 
$l \approx \zeta_{ACE}$ and $m \approx n \approx 1$. 
From this anisotropic expansion, and assuming a circular cross section for the cloud at ACE, we predict an oblate shape at Ulysses 
with an aspect ratio of the order of $(5.4)^{1-\zeta_{ACE}} \approx 1.55$.
If this anisotropic expansion is present also before the cloud reaches 1~AU, this aspect ratio could be even a bit larger.

\begin{table}
\caption{MHD quantities calculated according to each model (see Sect.~\ref{3.Magnetic}).
The first column indicates the model, the second shows the name of the global MHD 
quantities and their units, the next two columns show ACE and Ulysses results, 
and the last column shows the percentage of decay between ACE and Ulysses results.}
\label{table:Hflux}
\centering
         \begin{tabular}[h]{  l  l  c  c  c }
\hline
Model & Parameter & ACE & Ulysses & \% of decay \\
\hline
DM-in  & $F_z$/Mx $10^{21}$        & ~1.2  & ~0.8  & 33 \\
EFI     & $F_z$/Mx $10^{21}$        & ~1.0  & ~0.9  & 11 \\
\hline
DM-in  & $F_y$/Mx $10^{21}$        & ~2.7  & ~2.5  & ~~7 \\
EFI     & $F_y$/Mx $10^{21}$        & ~2.2  & ~1.9  & 14 \\
\hline
DM-in  &   $H$/Mx$^{2}$ $10^{42}$ & -6.5  & -3.9~  & 40 \\
EFI     &   $H$/Mx$^{2}$ $10^{42}$  & -2.~~ & -1.8~  & 10 \\
\hline
DM-in  &   $E$/~erg $10^{28}$      & 18~~~~&  4.~~ & 78 \\
EFI     &   $E$/~erg $10^{28}$     & 14.5  &  3.8  & 74 \\
\hline
\end{tabular}
\end{table}

\subsection{Magnetic fluxes and helicity} \label{4.fluxesH}
 
From fitting all the free parameters of the expanding model to the observations (method EFI), we compute the global magnetic quantities (Section~\ref{3.Invariants_L}, Eqs.~(\ref{FzL}-\ref{EL})). 
We also compute these quantities from the direct observations in the in-bound branch, using the direct method (see Section~\ref{3.Invariants_DM}, Eqs. ~\ref{Fz_acum}-~\ref{E_DM}). 
We use a length $L=2$ for ACE and $L=10.8$ for Ulysses, because the MC is still connected to the Sun when observed at 1 AU and at 5.4 AU. 
There is a good agreement between the magnetic fluxes and helicities found at ACE and at Ulysses, with a trend to find slightly lower values at Ulysses, i.e., a small decay of $\approx$ 7-40 $\%$, depending on the method to estimate them (Table~\ref{table:Hflux}).  
We recall that the EFI method uses the full MC observations, so it is less accurate because of the strongly perturbed out-bound branch.   

The magnetic energy is not an MHD invariant.  In fact, its decay, assuming a self-similar expansion with $l=m=n$, is predicted as $e(t)^{-1}$ 
(Eq.~(\ref{EL})).  For the MC studied here, we expect a decay with a factor $5.4^{-l}$. This factor is in the range $[0.19,0.31]$ for $l$ in the range $[0.7,1.]$.  
 In fact, the results of Section~\ref{4.Prediction} indicate that the expansion is anisotropic ($l \approx 0.7$, $m\approx n \approx 1.$). 
The computation of the energy decay with such anisotropic evolution would require a theoretical development which is outside the scope of this paper.  
Still, the energy decay is expected to be within the above range.  Since the anisotropy in the coefficients $l,m,n$ is relatively small, 
an approximation of the magnetic energy decay is obtained using a mean expansion of $(l+m+n)/3=0.9$, which implies an energy decay $\approx 0.22$. 
From the last two rows of Table~\ref{table:Hflux}, 
the observed decay between 1 AU and 5.4 AU is 0.22 and 0.26 for DM and EFI, respectively. 
There is an excellent agreement with the theoretically expected decay, even when we have simplified the analysis to $l=m=n$ and cylindrical symmetry. 

\section{Solar source of the MC} \label{sec:5}

\subsection{Searching for the solar source} \label{5.Finding}

The first step to determine the MC source on the Sun is to delimit the time
at which the solar event could have happened. We compute the approximate
transit time from Sun to Earth using the MC average velocity at ACE (see
Sect.~\ref{4.Expansion}). Considering that the cloud has travelled $1$~AU at a
constant velocity of $\Vc \approx 350$~km/s (where we neglect the
acceleration, which is important only in the first stages of the CME
ejection), we find $\tau \approx 1~{\rm AU}/\Vc \approx 5$
days. As the structure was observed by ACE starting on 4 March, we
search for solar ejective events that occurred 5 days before, around
28 February $\pm 1$ day.

\begin{figure}
\centering
\includegraphics[width=0.45\textwidth,clip=]{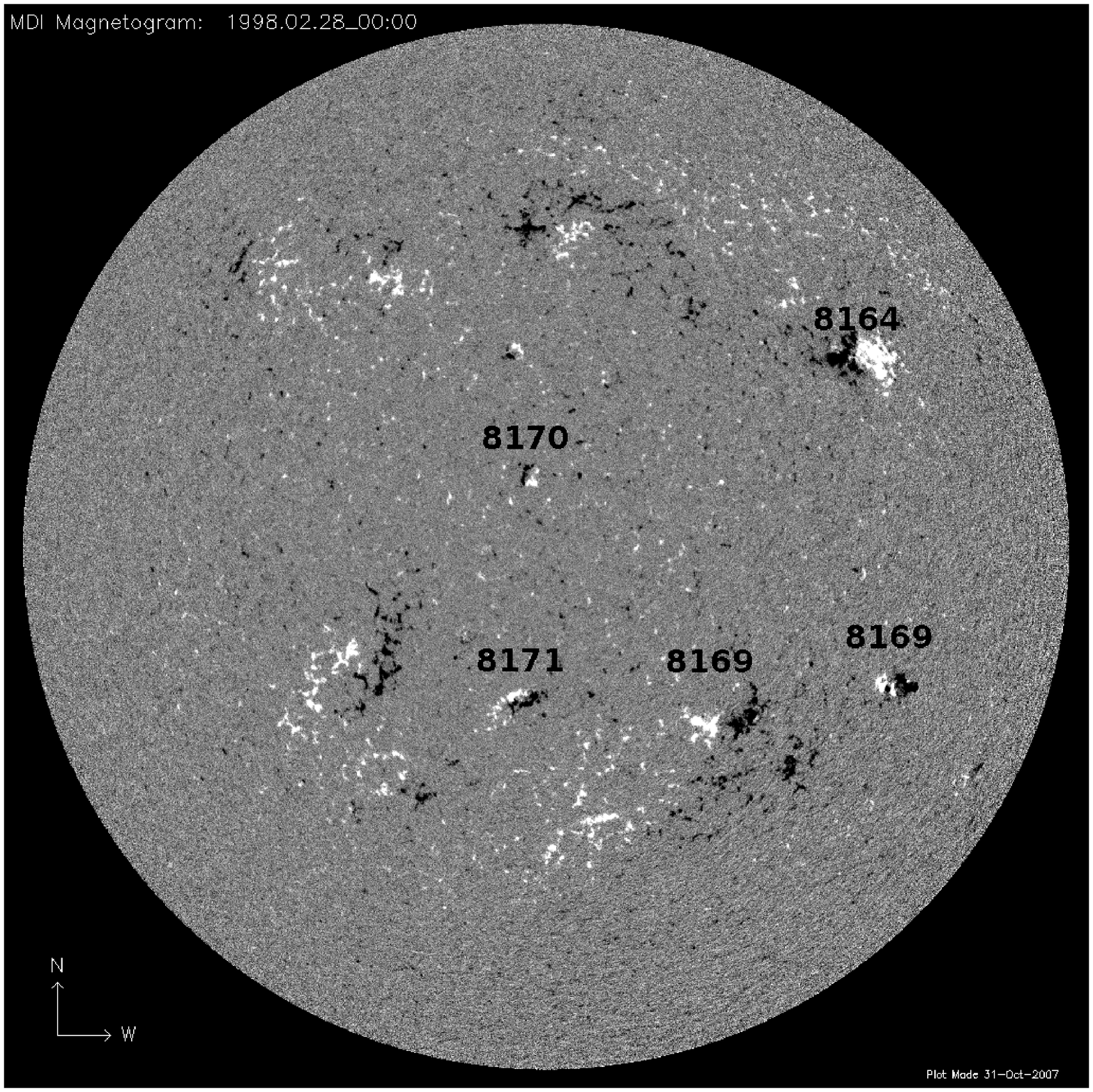}
\includegraphics[width=0.45\textwidth,clip=]{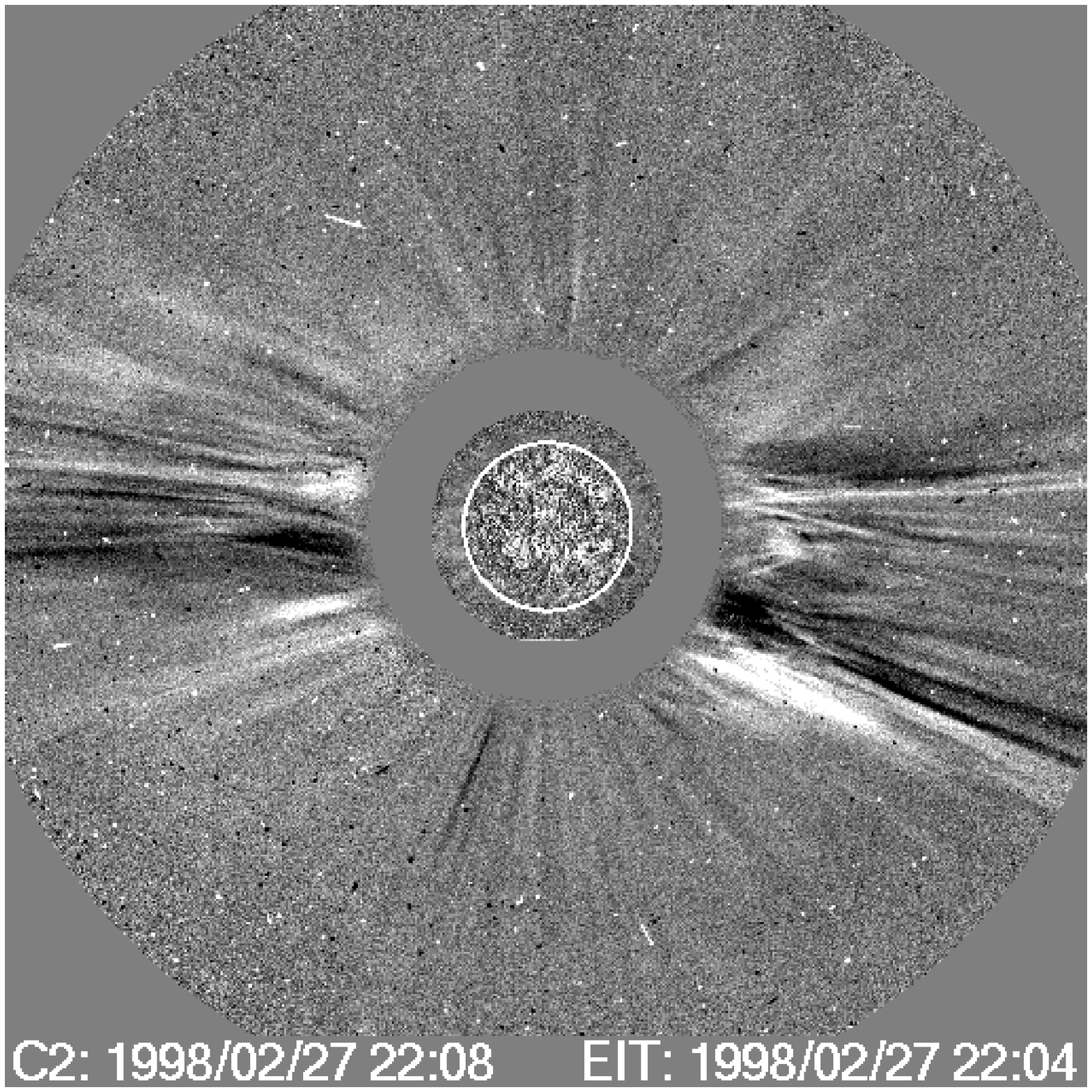}
\caption{(top) MDI full disk magnetogram on 28 February, 1998, at 00:00 UT
(positive/negative magnetic field polarities are indicated in
white/black colour).
The 5 ARs present on the solar disk are indicated with their NOAA numbers.
Notice the location of AR 8171 and AR 8164. The presence of magnetic tongues
with a shape compatible with a positive magnetic helicity sign is evident in
AR 8171. (bottom) LASCO C2 running difference image showing the halo CME at
22:08 UT together with the closest in time EIT running difference image in
195~\AA.}
\label{fig_MDI_LASCO}
\end{figure}

From 28 February to 1 March, 1998, 5 numbered active regions (ARs) were
present on the solar disk (see top panel of Fig.~\ref{fig_MDI_LASCO}). Only very low
X-ray class flares occurred in this period of time, most of them were class B (3 on 27 February, 2 on 28 February, 6 on 1 March) and 3 reached class C on 1 March (see the X-ray light curve from the Geostationary Operational Environmental Satellites
in http://www.solarmonitor.org and the corresponding list of events).
Two of the latter C-class events occurred in AR 8169 which was located at
$\approx$ S21W74 at the time of the flares. We have found no AR associated
to the observed B class flares.

We have also looked for the CMEs that occurred from 27 February to 1 March
in the catalogue of the Large Angle and Spectroscopic Coronagraph
\citep[LASCO,][]{Brueckner95} on board the Solar and Heliospheric
Observatory (SOHO, see http://cdaw.gsfc.nasa.gov/CME/list).
Most CMEs on those days had angular widths not larger than 70\degr\
and originated from the eastern solar limb, except for a halo and a partial
halo CME. 

The halo CME first appeared in LASCO C2 on 27 February at 20:07 UT.
This was a poor event inserted in the LASCO catalogue after a revision on
January 2006. The CME is clearly visible only in LASCO C2 running difference
images (see bottom panel of Fig.~\ref{fig_MDI_LASCO}), in particular after
$\approx$ 22:00 UT when its front has already left the C2 field of view 
(LASCO C2 field of view from 20:07 UT until 22:08 UT is only partial). 
The speed of the CME leading edge seen in LASCO/C2 and C3  
is $\approx 420$~km/s from a linear fit, while a fit with a second order polynomia provides 
$\approx 340$~km/s at 20 $R_{\sun}$; these values are in agreement with the velocity measured 
at the front of the MC at ACE ($\approx 385$~km/s), taking into account that the MC velocity 
is expected to be slightly modified by the interaction with the surrounding wind during its travel 
to 1~AU. Moreover, a velocity of $\approx 340$~km/s gives a travel time of $\approx 4.5$ days, so as expected.

On the other hand, the partial halo CME first appears in LASCO C2 on 28 February
at 12:48 UT. Its central position angle is 236\degr\ and its speed from a
second order fitting is $\approx 225$~km/s, which is too low considering the
MC arrival time at ACE. Furthermore, from an analysis of
images of the Extreme Ultraviolet Imaging Telescope
\citep[EIT][]{Delaboudiniere95} on board SOHO in 195~\AA~ the CME seems to be a
backside event. Therefore, the halo CME is the candidate to be the
MC solar counterpart.

To find the source of the halo CME on the Sun, we analyze EIT images
obtained on 27 February starting 2 hours before the halo CME appearance
in LASCO C2. EIT was working in CME watch mode at that time and, therefore, only images in 195~\AA~with half spatial resolution and with a temporal cadence of
$\approx$ 15 min, or larger, are available.
A sequence of 4 images with full spatial resolution in all EIT spectral
bands was taken at $\approx$ 07:00 UT, 13:00 UT, 19:00 UT. Furthermore,
there is an extended data gap in EIT starting at around 20:00 UT until around
22:00 UT. Considering that all events on 27 February were of very low
intensity (a B2.3 flare at $\approx$ 18:20 UT, a B1.5 flare
at $\approx$ 19:17 UT, and a B4.3 at $\approx$ 22:55 UT after the
halo CME), the low temporal cadence and spatial resolution
of EIT images, and the existence of the data gap, we have not been able to
unambiguously identify the CME source region. However, from these images and
those from the Soft X-ray Telescope \citep[SXT,][]{Tsuneta91}, it is evident that 2 of the five ARs present on the solar disk displayed signatures of activity (compare,
in particular, the image at 20:17 UT with the previous and following one in SXT AlMg movie in http://cdaw.gsfc.nasa.gov/CME\_/list/); these are AR 8171 and AR 8164
(see Fig.~\ref{fig_MDI_LASCO}).

From the two ARs that could be the source of the halo CME on 27 February
at 20:07 UT, AR 8171 was located at S24E02 which is an appropriate location
for a solar region to be the source of a cloud observed at Earth.
However, the magnetic flux in this AR
is $\approx$ 2.0 $\times$ 10$^{21}$ Mx. This value is small when compared to
the range for the axial MC flux measured at ACE
(1. - 1.2 $\times$ 10$^{21}$ Mx) since, in general,
a MC axial flux is 10\% of the AR magnetic flux \citep{Lepping97}.
Furthermore, according to the distribution of the photospheric field of the AR polarities
\citep[i.e., the shape of magnetic tongues, see][and Fig.~\ref{fig_MDI_LASCO}]{LopezFuentes00,Luoni11} the magnetic
field helicity in this AR is positive, which is opposite to the MC
magnetic helicity sign. 
Finally, the leading polarity of AR 8171 is negative, and with a positive helicity this implies that the magnetic field component along the polarity inversion line (PIL) points
from solar east to west. Since the magnetic field component along the PIL
is related to the axial MC field component, this is not compatible with the MC axial field orientation at ACE that points to the solar east (Figure~\ref{fig_GSE_ACE}).
 Therefore, we conclude that AR 8171 cannot be the
solar source region of the halo CME despite its appropriate location on the
disk.

The other possible CME source region is AR 8164, located at N16W32 on
27 February at $\approx$ 20:00 UT. This region is far from the central
meridian, considering this location and a radial ejection, one would
expect that ACE would have crossed the ejected flux rope eastern leg;
however, ACE crossed the MC front (see Section~\ref{4.MC_ACE})
which implies that during the ejection the flux rope suffered a deflection towards the east; this probably 
occurred low in the corona as the CME is a halo.
Concerning the AR magnetic flux, its value is
$\approx$ 10. $\times$ 10$^{21}$ Mx, which is high enough to explain
the MC axial magnetic flux as we discussed previously.

\begin{figure}
\resizebox{\hsize}{!}{\includegraphics[width=4cm]{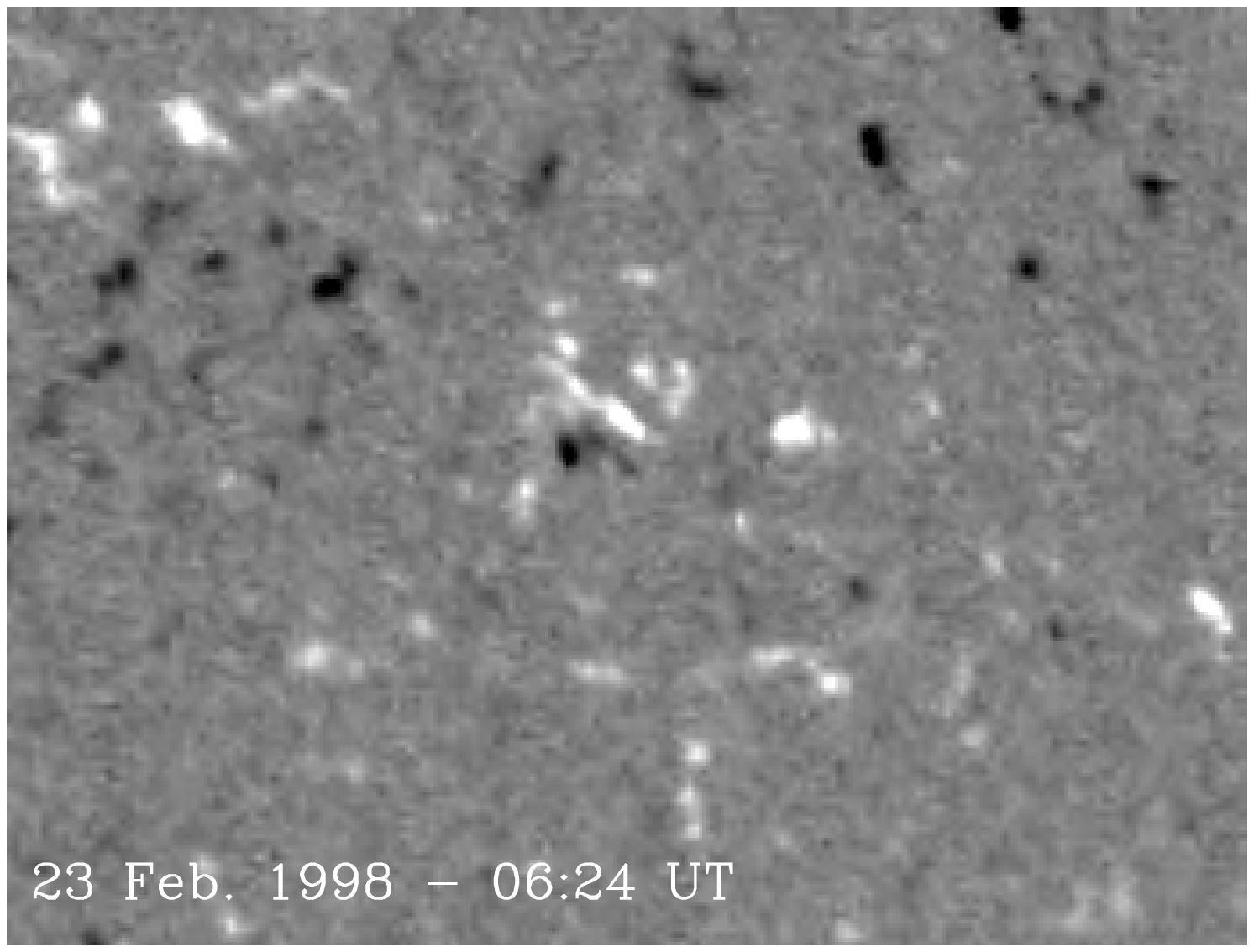} 
                      \includegraphics[width=4cm]{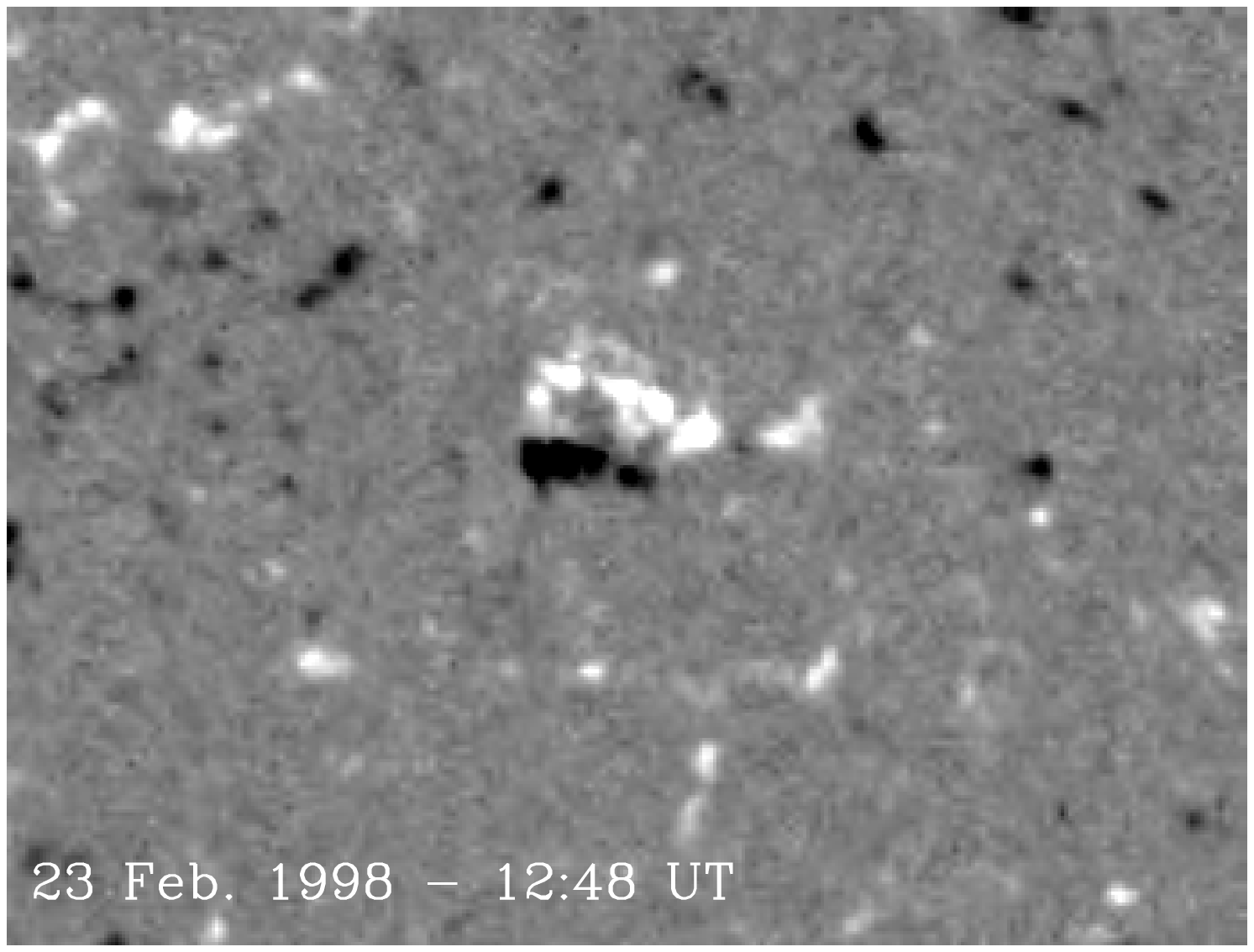}}
\resizebox{\hsize}{!}{\includegraphics[width=4cm]{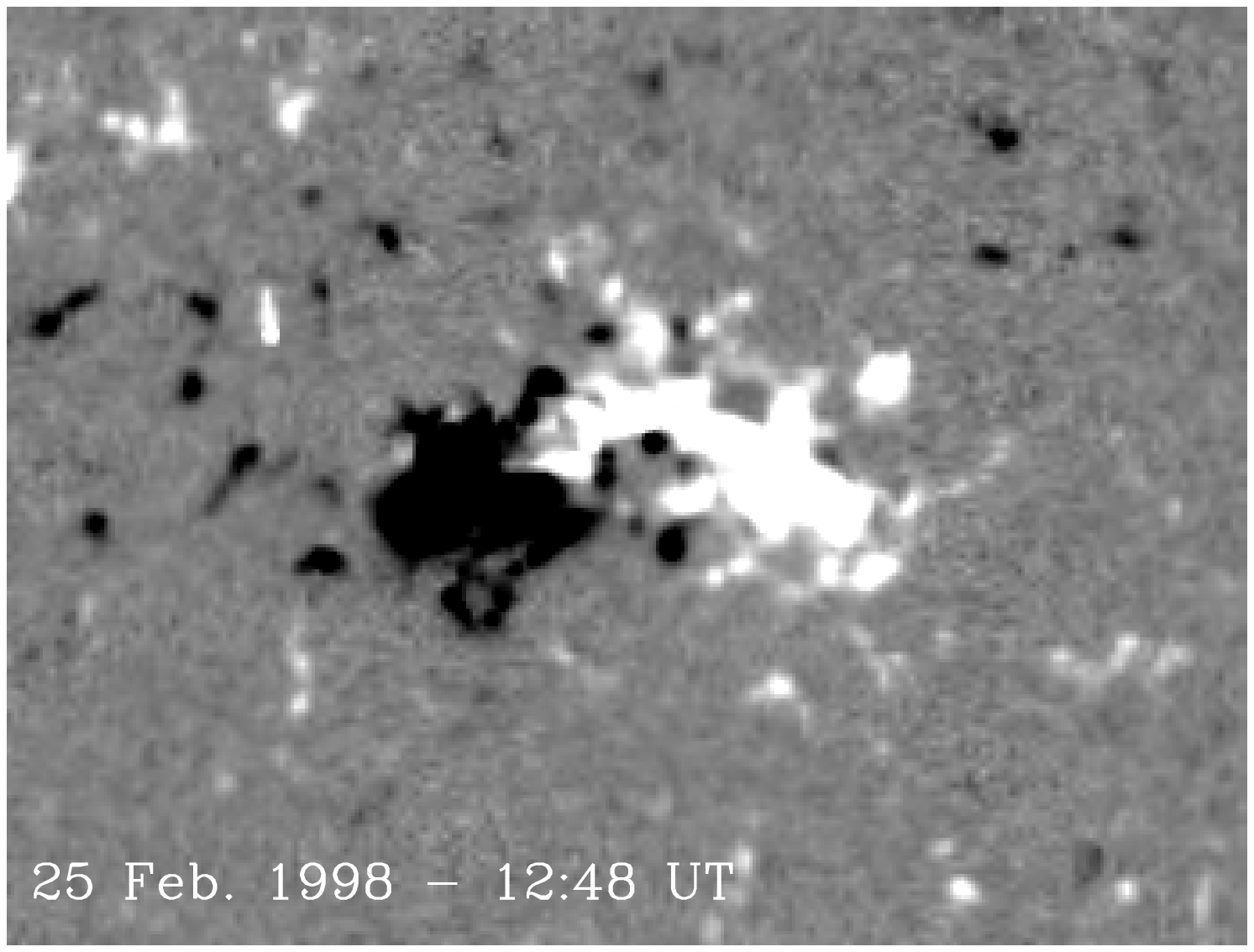}
                      \includegraphics[width=4cm]{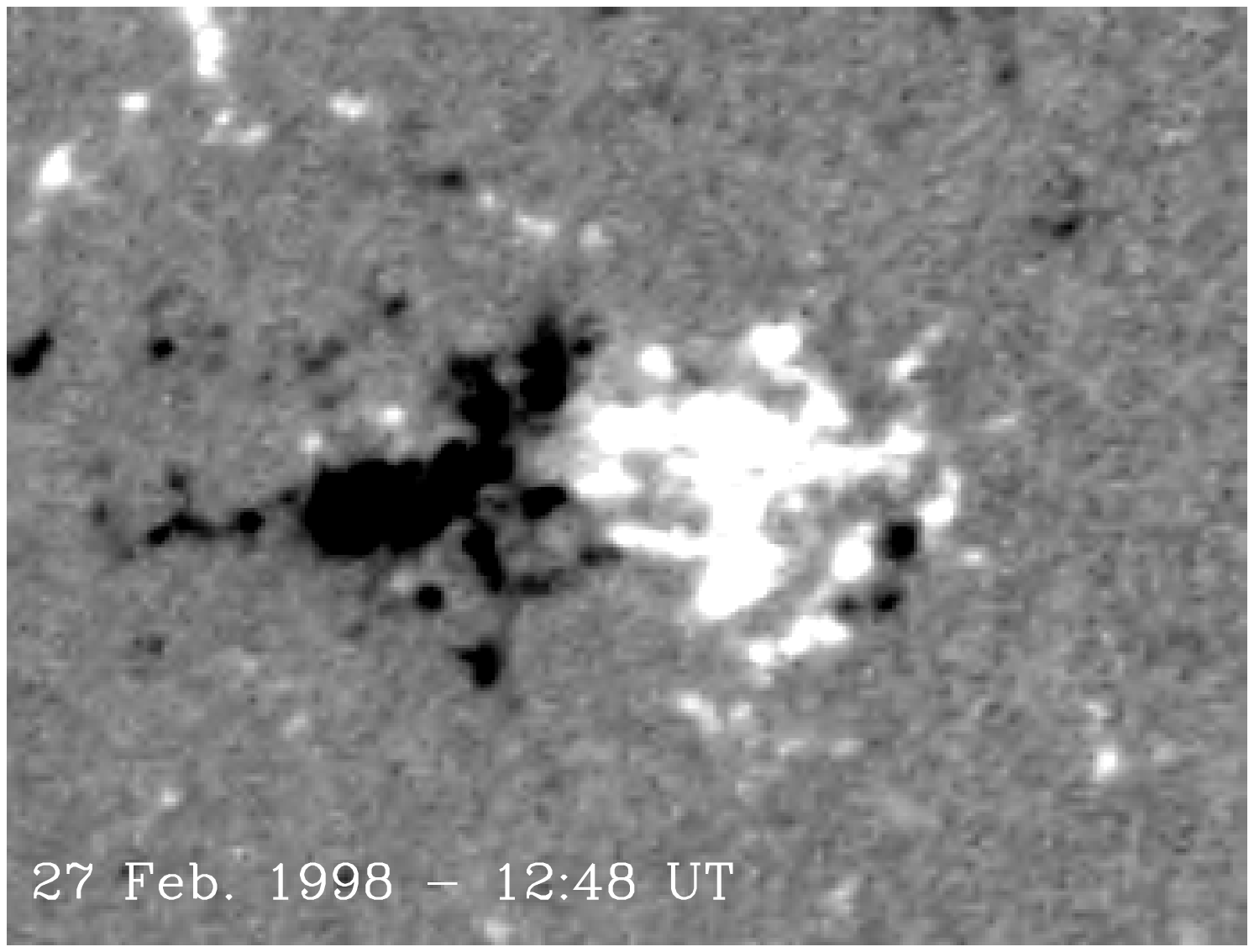}}
\caption{MDI magnetograms of AR 8164 showing the photospheric evolution
of its main polarities. The shape and evolution of the magnetic tongues
indicates that the magnetic helicity of AR 8164 is negative. The magnetograms
have been rotated to the central meridian position of the AR
(positive/negative magnetic field polarities are indicated in
white/black color). The size of the field of view is the same
in all panels.}
\label{fig_MDI}
\end{figure}

The magnetic helicity sign of AR 8164 is negative, as shown by the
shape and evolution of its photospheric polarities in Figure~\ref{fig_MDI}
(the spatial organization of the magnetic tongues on 23-25 February).
This sign is also confirmed by the coronal field model of the region
(see Sect~\ref{5.Physical}) and agrees with the MC helicity sign.

Conversely to AR 8171, the leading polarity in AR 8164 is positive, implying
a magnetic field component pointing from west to east along the PIL. In this
case, this direction is compatible with the orientation of MC axial field.
Furthermore, the PIL forms an angle of $\approx 60\degr $ in the clockwise direction
with the solar equator, while the MC axis lies almost on the ecliptic
(see Sect.~\ref{4.MC_ACE}).
This difference between the PIL on the Sun and the MC axial direction can be
explained by a counter-clockwise rotation of the ejected flux
rope, as expected, since its helicity is negative \citep{Torok05,Green07}.

From the previous analysis, we conclude that AR 8164 is the most plausible
source of the halo CME on 27 February, 1998, which can be
the counterpart of the MC observed at ACE on 4-5 March. In the next
section we compute the magnetic helicity of the AR before and after the
ejection and its variation; this value is used as a proxy of
the magnetic helicity carried away from the Sun by the CME.

\begin{figure}
\includegraphics[width=0.5\textwidth,clip=]{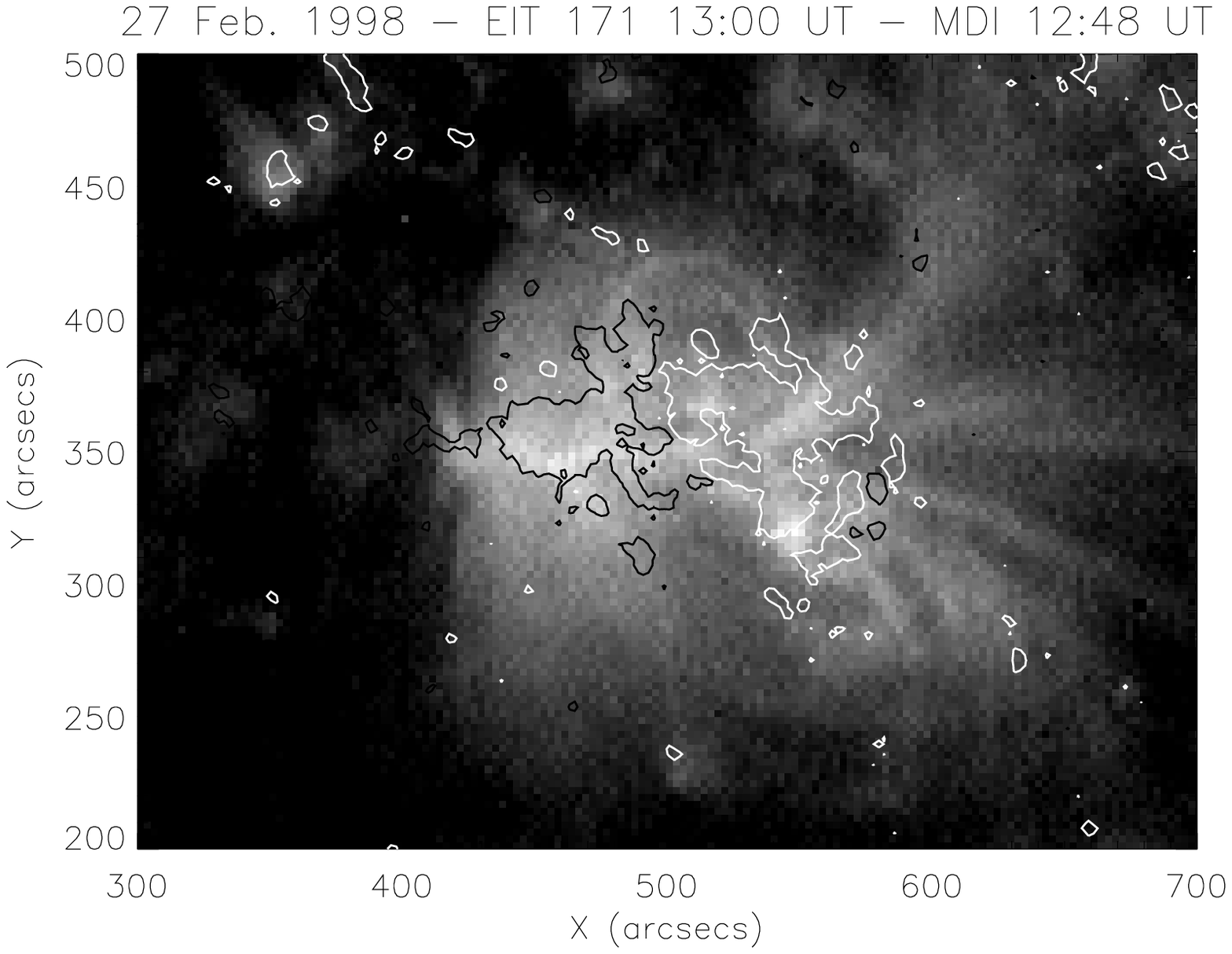}
\includegraphics[width=0.49\textwidth,clip=]{27feb-1300-015_limpias_bw}
\caption{(top) EIT image in 171~\AA\ at 13:00 UT on 27 February, 1998, with
two MDI isocontours overlaid ($\pm 100$ G) shown with continuous lines
(positive: white line, negative: black line). The MDI magnetogram
was taken at 12:48 UT. (bottom) The same EIT image with more isocontours
($\pm$ 100, 500 G) shown with grey lines and computed field lines
superimposed (black continuous lines).
Both axes are measured in Mm in the local solar frame.}
  \label{fig_model}
\end{figure}

\subsection{Physical properties of the solar source} \label{5.Physical}

  Using AR 8164 MDI magnetograms, we have extrapolated the
observed photospheric line of sight component of the field to the
corona under the linear (or constant $\alpha$) force-free field
assumption: $\vec{\nabla} \times \vec{B}= \alpha \vec{B}$.  We have
used a fast Fourier transform method as proposed by
\citet{Alissandrakis81} and the transformation of coordinates discussed
in \citet{Demoulin97}.  The value of $\alpha$ is chosen so as to best fit
the observed coronal loops at a given time. We need high spatial
resolution images to identify independent loops, which are not available
at times close enough before the ejection. Then, we have used the full spatial resolution images obtained by EIT in 171~\AA~
on 27 February at 13:00 UT; this is the closest time to the event in
which coronal loops are visible. The boundary conditions for the model
are given by the MDI magnetogram at 12:48 UT on the same day.
The value of $\alpha$ is determined through
an iterative process that has been explained in \citet{Green02}.
The value of $\alpha$ that best fits the observed loops is
$\alpha$ = -9.4$\times 10^{-2}$~Mm$^{-1}$ (Figure~\ref{fig_model}). 

Once the coronal model is determined, we compute the relative coronal
magnetic helicity, $H_{cor}$, following \citet{Berger85}. In particular,
we use a linearized  version of the expression given by
\citet[~see his Eq.~A23]{Berger85} as has been done in previous works by
\citet{Mandrini05} and \citet{Luoni05}. Following this approach, the
magnetic helicity content in the coronal field before the ejection
is $H_{cor}$ = -11.4 $\times$ 10$^{42}$~Mx$^2$.

When a flux rope is ejected from the Sun into the
IP medium, it carries part of the magnetic helicity
contained in the coronal field.  Therefore, we have to compute the
variation of the coronal magnetic helicity, by subtracting its
value before and after an eruptive event, to compare this
quantity to the corresponding one in the associated IP event. As
done before, we search for an EIT image in 171~\AA~ with full spatial resolution after the ejection in which loops could be visible. However, in this
case, the AR is much closer to the east limb and projection effects, added
to the low intensity of the coronal structures, make it even more difficult
to distinguish the shape of individual loops; as a result, $\alpha$ cannot
be unambiguously determined, i.e. we can adjust the global shape of EIT
brightness with more than one $\alpha$ value. We selected the EIT image
at 01:20 UT on 28 February and, following a conservative approach, we
determine a lower bound for the coronal magnetic helicity variation.
We select the closest in time MDI magnetogram
(at 01:36 UT on 28 February) and, using the previously determined value
for $\alpha$, we compute $H_{cor}$. As the AR magnetic field is decaying, its
flux is lower than before the CME ($\approx$ 7.0 $\times$ 10$^{21}$ Mx);
therefore, $H_{cor}$ is also lower,  $H_{cor}$ = -8.1 $\times$ 10$^{42}$~Mx$^2$.
The real value of the coronal magnetic helicity after the CME
should be even lower than the later one, as we expect that the field
relaxes to a closer to potential state. To determine the range of variation
for $H_{cor}$, we also compute its value taking the lowest $\alpha$ value
($\alpha$ = -6.3 $\times 10^{-2}$~Mm$^{-1}$) that still gives a good fitting to
the global shape of EIT brightness after the CME; in this case,
$H_{cor}$ = -5.4 $\times$ 10$^{42}$~Mx$^2$. Considering the two
values determined for $H_{cor}$ after the CME, we estimate that
$3.3 \times 10^{42}$~Mx$^2$ $\leq |\Delta H_{cor}| \leq 6.0 \times
10^{42}$~Mx$^2$.

\subsection{Link with the observed MC} \label{5.Link}

From estimations of the helicity content when the cloud was observed at ACE and Ulysses, using the EFI and DM methods (see Table~\ref{table:Hflux}), we found 
$2 \times 10^{42}$~Mx$^2 \leq |H_{MC}| \leq 6 \times 10^{42}$~Mx$^2$, 
which is fully consistent with the range  
found for the release of magnetic helicity in the corona during the CME eruption.

A fraction of the total magnetic flux of AR 8164 ($\approx 10 \times 10^{21}$~Mx) 
is enough to account for the magnetic flux in the MC at ACE,
$F_z$ ($\approx 10^{21}$~Mx) +  $F_y$ ($\approx 2 \times 10^{21}$~Mx) $\approx  3 \times 10^{21}$~Mx.

Thus, we have found qualitative and quantitative proofs that let us 
associate the halo CME observed by LASCO C2 on 27 February, 1998, to its solar
source region (AR 8164) and to its interplanetary counterpart, the MC
observed at ACE on 4-5 March, 1998, and at Ulysses on 24-28 March, 1998.

\section{Summary and Conclusions}\label{sec:6}

We have studied a magnetic cloud which was observed \insitu\ by two spacecraft 
(ACE and Ulysses) in an almost radial alignment with the Sun ($\approx 2\degr $ for latitude and $\approx 6\degr $ for longitude) 
and significantly separated in distance (ACE at 1 AU and Ulysses at 5.4 AU). 
This is an uncommon geometrical scenario and it is very appropriate for multi-spacecraft analysis of MC evolution. 
In each of the spacecraft locations, we have analyzed the cloud in the local frame 
(attached to the flux-rope axis) and quantified magnetic fluxes, helicity, and energy, 
using an expansion model of an initial Lundquist field (EFI) and a direct method (DM), which permits the computation of global magnetic quantities directly from the observed magnetic field 
\citep{Dasso05b}.
We also computed the local non-dimensional expansion rate ($\zeta$) at ACE and at Ulysses from the observed bulk velocity profiles \citep[as defined in][]{Demoulin08}.

We found close values of the normalized expansion rate along the solar radial direction, $\zeta_{ACE} = 0.74$ and $\zeta_{Uly} = 0.67$, 
as measured from the radial proton velocity at ACE and Ulysses, respectively. 
From the measured $\zeta_{ACE}$ in the radial direction and assuming a self-similar expansion proportional to the solar distance in the 
ortho-radial directions, we successfully predicted  at Ulysses the values of the MC size, 
the mean values for density and magnetic field components from the values of these quantities measured at ACE.

Next, comparing observations at Ulysses with different models for anisotropic expansions in the two directions that cannot be directly observed ($m$ and $n$), we found that the expansion on the plane perpendicular to the cloud axis is larger than in the direction perpendicular to the radial direction from the Sun. 
Based on the quantification of this anisotropic expansion, 
we conclude that the initial isotropic structure at 1AU will develop an oblate shape such that its aspect ratio would be $\approx 1.6$ at 5.4 AU (i.e., the major axis $\approx 60 \%$ larger than the minor one, with the major one perpendicular to the radial direction 
from the Sun).

From a comparison of the transit time, axis orientation, magnetic fluxes, 
and magnetic helicity, and considering all the solar sources inside
a time window, we have also identified the possible source at the Sun for this event, finding an agreement between the amounts of magnetic fluxes and helicity, 
in consistence with a rough conservation of these so-called ideal-MHD invariants.

In particular, we found that there is a small decay of the magnetic fluxes and helicity between 1 and 5.4 AU, with a $\approx 10 \%$ of decay for $F_z$ and $F_y$, and a decay of $\approx 10 \%$ for the magnetic helicity when the EFI method is used and $\approx 40 \%$ when DM is used, respectively.
These decays can be due to a possible erosion or pealing of the flux rope during its travel, for instance because of 
magnetic reconnection with the surrounding SW \citep[e.g., ][]{Dasso06}.

For a self-similar expansion and known expansion rates, it is possible to theoretically derive the decay of the magnetic energy during the travel of the flux rope in the SW.  From the observed values of $\zeta$ and modeling the expansion rates in the other two directions ($m$ and $n$), we predict its decay during the travel from 1 AU to 5.4 AU. The measurements confirm this expected magnetic energy decay (from $(15-18) \times 10^{28}$ erg to $4 \times 10^{28}$ erg).

Summarizing, in this work we validate for the first time that the local expansion rate ($\zeta$) observed from the velocity profile can be used to make predictions of the decay of mass density and magnetic quantities. 
From the comparison of detailed predictions and observations of the decays of these quantities, we provide empirical evidence
about the quantification of the anisotropic expansion of magnetic clouds beyond Earth, to 5 AUs. Finally, we quantify how much the 
so-called ideal-MHD invariants are conserved in flux ropes traveling in the solar wind. Then, this kind of combined studies, using multi-spacecraft techniques, is a powerful approach to improve our knowledge of the properties and evolution of magnetized plasma structures ejected from the Sun.


\begin{figure*} 
\centerline{
\includegraphics[width=8cm, clip=]{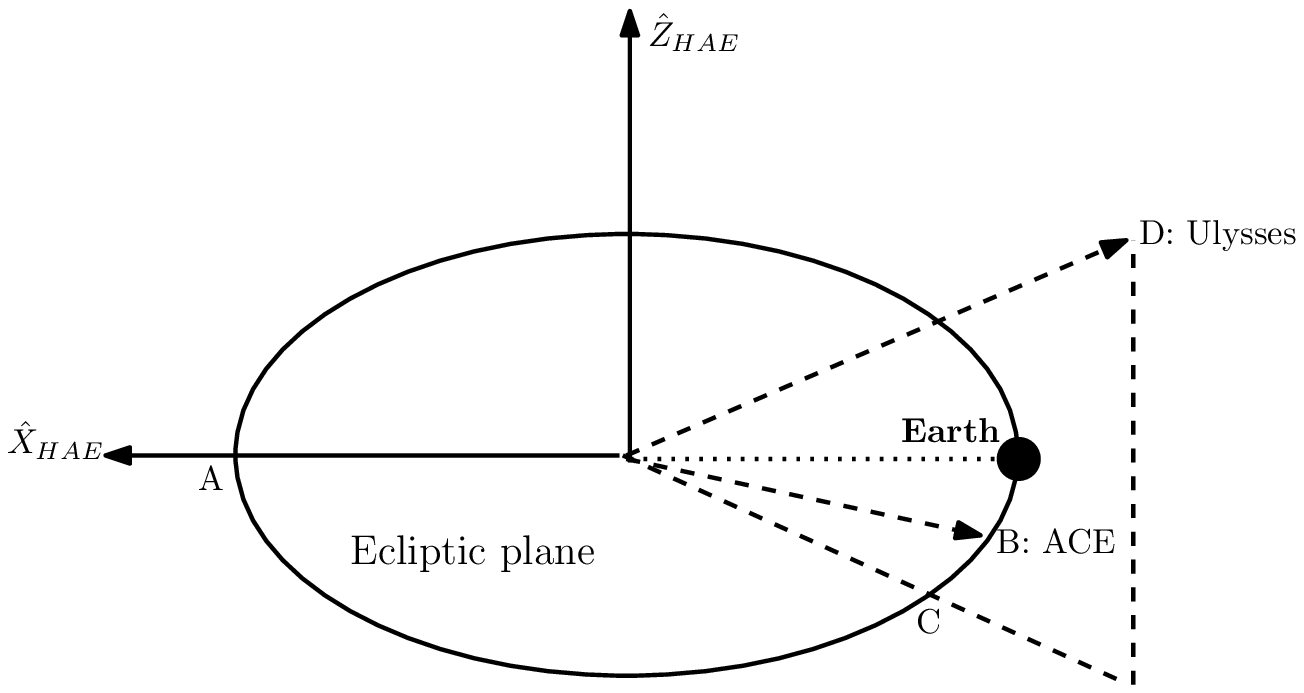}
\qquad
\includegraphics[width=8cm, clip=]{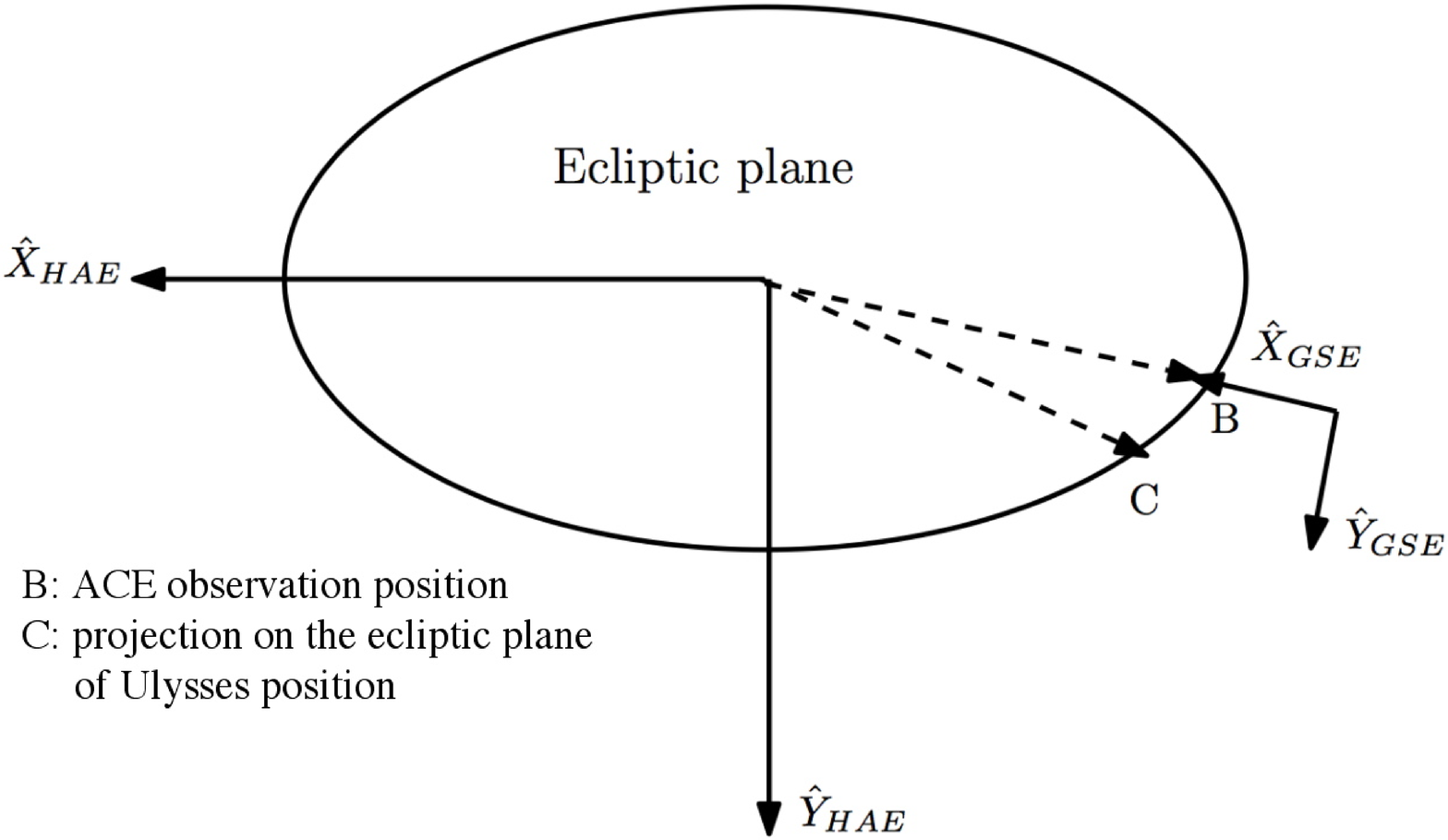}}
\caption{Schema of the positions of each spacecraft at the 
observed times (left). Schema showing the relationship between the GSE 
and the HAE system of coordinates (right).}
\label{fig_schema}
\end{figure*}

\begin{appendix}
\section{Transformation of coordinates between ACE and Ulysses} \label{sec:appendix} 

We describe below the transformation of coordinates between the natural system where
the vector data of Ulysses are provided ($\uvec{R},\uvec{T},\uvec{N}$) to the GSE system at the time when the cloud was observed by ACE. Then, we provide a common frame to compare vector observations made at Ulysses and ACE (as shown in Figures~\ref{fig_cloud_ACE}-~\ref{fig_cloud_Uly}).

A third coordinate system, the HAE system (Heliocentric Aries Ecliptic, \citet{Fraenz02}), is used since the location of both spacecraft are known in the HAE system. 
In this frame, $\uvec{Z}_{HAE}$ is normal to the ecliptic plane, and $\uvec{X}_{HAE}$ is positive towards the first point of Aries (from Earth to Sun at the vernal equinox, $\approx 21$ March). 
At the moment in which the cloud was observed by ACE, it was located at a longitude 
$\epsilon_{A}=163\degr $ and a latitude $\tau_A=0\degr $ in the HAE system.
When the MC was observed by Ulysses, this spacecraft was located at a longitude of $\epsilon_U=157\degr $ and at a latitude of $\tau_U=2\degr\ $. 

The solar equator is defined as the plane normal to the Sun's rotation vector ($\vec{\Omega}$) and it is inclined by $\alpha \approx 7.25\degr $ from $\uvec{Z}_{HAE}$.
In 2000, the solar equator plane intersected the ecliptic plane at a HAE longitude 
of $\approx 75.6\degr $. Then, the angle between the projection of $\vec{\Omega}$ on
the ecliptic is $\beta \approx 14.4\degr $.
When we write $\uvec{R}$ and $\vec{\Omega}$ in the HAE system of coordinates, we obtain:

   \begin{eqnarray}
       \uvec{R} &=& \cos(\tau_U) \cos(\epsilon_U)~\uvec{X}_{HAE}
                +  \cos(\tau_U) \sin(\epsilon_U)~\uvec{Y}_{HAE} \nonumber \\
               &+& \sin(\tau_U)                 ~\uvec{Z}_{HAE}   \,,\\
   \vec{\Omega}&=& \sin(\alpha) \cos(\beta)~\uvec{X}_{HAE}
                -  \sin(\alpha) \sin(\beta)~\uvec{Y}_{HAE} \nonumber \\
               &+& \cos(\alpha)            ~\uvec{Z}_{HAE}   \,.
\end{eqnarray}
To obtain $\uvec{T}$ and $\uvec{N}$ at Ulysses in HAE coordinates, we compute:
   \begin{eqnarray}
   \uvec{T}&=&\frac{\vec{\Omega}\times\uvec{R}}{|\vec{\Omega}\times\uvec{R}|}\\
   \uvec{N}&=&\uvec{R}\times\uvec{T}   \,.
   \end{eqnarray} 
To go to the local GSE system of reference we do a last rotation around 
$\uvec{Z}_{HAE}$, which coincides with $\uvec{Z}_{GSE}$, in an angle $\delta=180\degr -\epsilon_A$ 
(see Fig.~\ref{fig_schema}).
\end{appendix}

\begin{acknowledgements}
The authors thank the referee for helpful comments which improved the clarity of the paper.
This research has made use of NASA's Space Physics Data Facility (SPDF).
We thank the ACE and Ulysses instrument teams. 
We thank their respective principal investigators, namely N.F. Ness (ACE/MAG), 
D.J. McComas (ACE/SWEPAM and Ulysses/SWOOPS), and A. Balogh (Ulysses/VHM).
The authors acknowledge financial support from ECOS-Sud
through their cooperative science program (N$^o$ A08U01).
This work was partially supported by the Argentinean grants:
UBACyT 20020090100264, UBACyT X127, PIP-2009-00825, PIP-2009-00166 (CONICET) and PICT-2007-00856 and PICT-2007-1790 (ANPCyT).
MSN acknowledges Funda\c c\~ao de Amparo a Pesquisa do Estado de S\~ao Paulo (FAPESP -- Brazil) for partial financial support.
SD and CHM are members of the Carrera del Investigador Cien\-t\'\i fi\-co (CIC), CONICET.
S Dasso acknowledges support from the Abdus Salam International Centre for Theoretical Physics (ICTP),
as provided in the frame of his regular associateship.
\end{acknowledgements}

\bibliographystyle{aa} 
\bibliography{Du-sole} 
\IfFileExists{\jobname.bbl}{}
{\typeout{}
\typeout{****************************************************}
\typeout{****************************************************}
\typeout{** Please run "bibtex \jobname" to optain}
\typeout{** the bibliography and then re-run LaTeX}
\typeout{** twice to fix the references!}
\typeout{****************************************************}
\typeout{****************************************************}
\typeout{}
}

\end{document}